# GW170817: Observation of Gravitational Waves from a Binary Neutron Star Inspiral

B. P. Abbott *et al.*[*]
(LIGO Scientific Collaboration and Virgo Collaboration)

On August 17, 2017 at 12:41:04 UTC the Advanced LIGO and Advanced Virgo gravitational-wave detectors made their first observation of a binary neutron star inspiral. The signal, GW170817, was detected with a combined signal-to-noise ratio of 32.4 and a false-alarm-rate estimate of less than one per $8.0 \times 10^4$ years. We infer the component masses of the binary to be between 0.86 and 2.26 $M_\odot$, in agreement with masses of known neutron stars. Restricting the component spins to the range inferred in binary neutron stars, we find the component masses to be in the range 1.17–1.60 $M_\odot$, with the total mass of the system $2.74^{+0.04}_{-0.01} M_\odot$. The source was localized within a sky region of 28 deg$^2$ (90% probability) and had a luminosity distance of $40^{+8}_{-14}$ Mpc, the closest and most precisely localized gravitational-wave signal yet. The association with the $\gamma$-ray burst GRB 170817A, detected by Fermi-GBM 1.7 s after the coalescence, corroborates the hypothesis of a neutron star merger and provides the first direct evidence of a link between these mergers and short $\gamma$-ray bursts. Subsequent identification of transient counterparts across the electromagnetic spectrum in the same location further supports the interpretation of this event as a neutron star merger. This unprecedented joint gravitational and electromagnetic observation provides insight into astrophysics, dense matter, gravitation, and cosmology.

## I. INTRODUCTION

On August 17, 2017, the LIGO-Virgo detector network observed a gravitational-wave signal from the inspiral of two low-mass compact objects consistent with a binary neutron star (BNS) merger. This discovery comes four decades after Hulse and Taylor discovered the first neutron star binary, PSR B1913+16 [1]. Observations of PSR B1913+16 found that its orbit was losing energy due to the emission of gravitational waves, providing the first indirect evidence of their existence [2]. As the orbit of a BNS system shrinks, the gravitational-wave luminosity increases, accelerating the inspiral. This process has long been predicted to produce a gravitational-wave signal observable by ground-based detectors [3–6] in the final minutes before the stars collide [7].

Since the Hulse-Taylor discovery, radio pulsar surveys have found several more BNS systems in our galaxy [8]. Understanding the orbital dynamics of these systems inspired detailed theoretical predictions for gravitational-wave signals from compact binaries [9–13]. Models of the population of compact binaries, informed by the known binary pulsars, predicted that the network of advanced gravitational-wave detectors operating at design sensitivity will observe between one BNS merger every few years to hundreds per year [14–21]. This detector network currently includes three Fabry-Perot-Michelson interferometers that measure spacetime strain induced by passing gravitational waves as a varying phase difference between laser light propagating in perpendicular arms: the two Advanced LIGO detectors (Hanford, WA and Livingston, LA) [22] and the Advanced Virgo detector (Cascina, Italy) [23].

Advanced LIGO's first observing run (O1), from September 12, 2015, to January 19, 2016, obtained 49 days of simultaneous observation time in two detectors. While two confirmed binary black hole (BBH) mergers were discovered [24–26], no detections or significant candidates had component masses lower than $5M_\odot$, placing a 90% credible upper limit of 12 600 Gpc$^{-3}$ yr$^{-1}$ on the rate of BNS mergers [27] (credible intervals throughout this Letter contain 90% of the posterior probability unless noted otherwise). This measurement did not impinge on the range of astrophysical predictions, which allow rates as high as ~10 000 Gpc$^{-3}$ yr$^{-1}$ [19].

The second observing run (O2) of Advanced LIGO, from November 30, 2016 to August 25, 2017, collected 117 days of simultaneous LIGO-detector observing time. Advanced Virgo joined the O2 run on August 1, 2017. At the time of this publication, two BBH detections have been announced [28,29] from the O2 run, and analysis is still in progress.

Toward the end of the O2 run a BNS signal, GW170817, was identified by matched filtering [7,30–33] the data against post-Newtonian waveform models [34–37]. This gravitational-wave signal is the loudest yet observed, with a combined signal-to-noise ratio (SNR) of 32.4 [38]. After

[*]Full author list given at the end of the Letter.

~100 s (calculated starting from 24 Hz) in the detectors' sensitive band, the inspiral signal ended at 12:41:04.4 UTC. In addition, a $\gamma$-ray burst was observed 1.7 s after the coalescence time [39–45]. The combination of data from the LIGO and Virgo detectors allowed a precise sky position localization to an area of 28 deg$^2$. This measurement enabled an electromagnetic follow-up campaign that identified a counterpart near the galaxy NGC 4993, consistent with the localization and distance inferred from gravitational-wave data [46–50].

From the gravitational-wave signal, the best measured combination of the masses is the chirp mass [51] $\mathcal{M} = 1.188^{+0.004}_{-0.002} M_\odot$. From the union of 90% credible intervals obtained using different waveform models (see Sec. IV for details), the total mass of the system is between 2.73 and 3.29 $M_\odot$. The individual masses are in the broad range of 0.86 to 2.26 $M_\odot$, due to correlations between their uncertainties. This suggests a BNS as the source of the gravitational-wave signal, as the total masses of known BNS systems are between 2.57 and 2.88 $M_\odot$ with components between 1.17 and ~1.6 $M_\odot$ [52]. Neutron stars in general have precisely measured masses as large as $2.01 \pm 0.04$ $M_\odot$ [53], whereas stellar-mass black holes found in binaries in our galaxy have masses substantially greater than the components of GW170817 [54–56].

Gravitational-wave observations alone are able to measure the masses of the two objects and set a lower limit on their compactness, but the results presented here do not exclude objects more compact than neutron stars such as quark stars, black holes, or more exotic objects [57–61]. The detection of GRB 170817A and subsequent electromagnetic emission demonstrates the presence of matter. Moreover, although a neutron star–black hole system is not ruled out, the consistency of the mass estimates with the dynamically measured masses of known neutron stars in binaries, and their inconsistency with the masses of known black holes in galactic binary systems, suggests the source was composed of two neutron stars.

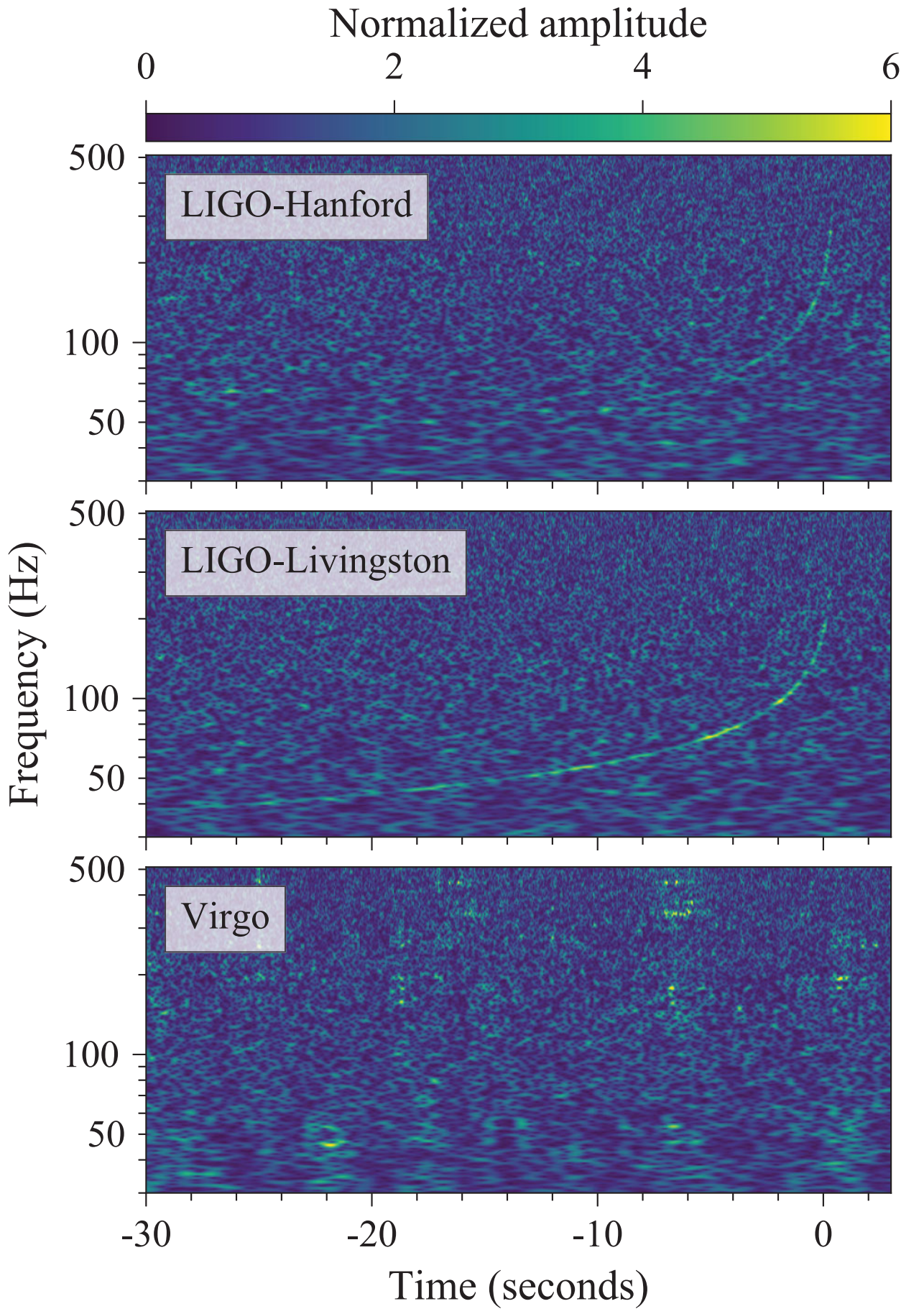

FIG. 1. Time-frequency representations [65] of data containing the gravitational-wave event GW170817, observed by the LIGO-Hanford (top), LIGO-Livingston (middle), and Virgo (bottom) detectors. Times are shown relative to August 17, 2017 12:41:04 UTC. The amplitude scale in each detector is normalized to that detector's noise amplitude spectral density. In the LIGO data, independently observable noise sources and a glitch that occurred in the LIGO-Livingston detector have been subtracted, as described in the text. This noise mitigation is the same as that used for the results presented in Sec. IV.

## II. DATA

At the time of GW170817, the Advanced LIGO detectors and the Advanced Virgo detector were in observing mode. The maximum distances at which the LIGO-Livingston and LIGO-Hanford detectors could detect a BNS system (SNR = 8), known as the detector horizon [32,62,63], were 218 Mpc and 107 Mpc, while for Virgo the horizon was 58 Mpc. The GEO600 detector [64] was also operating at the time, but its sensitivity was insufficient to contribute to the analysis of the inspiral. The configuration of the detectors at the time of GW170817 is summarized in [29].

A time-frequency representation [65] of the data from all three detectors around the time of the signal is shown in Fig 1. The signal is clearly visible in the LIGO-Hanford and LIGO-Livingston data. The signal is not visible in the Virgo data due to the lower BNS horizon and the direction of the source with respect to the detector's antenna pattern.

Figure 1 illustrates the data as they were analyzed to determine astrophysical source properties. After data collection, several independently measured terrestrial contributions to the detector noise were subtracted from the LIGO data using Wiener filtering [66], as described in [67–70]. This subtraction removed calibration lines and 60 Hz ac power mains harmonics from both LIGO data streams. The sensitivity of the LIGO-Hanford detector was particularly improved by the subtraction of laser pointing noise; several broad peaks in the 150–800 Hz region were effectively removed, increasing the BNS horizon of that detector by 26%.

Additionally, a short instrumental noise transient appeared in the LIGO-Livingston detector 1.1 s before the coalescence time of GW170817 as shown in Fig. 2. This transient noise, or glitch [71], produced a very brief (less than 5 ms) saturation in the digital-to-analog converter of the feedback signal controlling the position of the test masses. Similar glitches are registered roughly once every few hours in each of the LIGO detectors with no temporal correlation between the LIGO sites. Their cause remains unknown. To mitigate the effect on the results presented in Sec. III, the search analyses applied a window function to zero out the data around the glitch [72,73], following the treatment of other high-amplitude glitches used in the O1 analysis [74]. To accurately determine the properties of GW170817 (as reported in Sec. IV) in addition to the noise subtraction described above, the glitch was modeled with a time-frequency wavelet reconstruction [75] and subtracted from the data, as shown in Fig. 2.

Following the procedures developed for prior gravitational-wave detections [29,78], we conclude there is no environmental disturbance observed by LIGO environmental sensors [79] that could account for the GW170817 signal.

The Virgo data, used for sky localization and an estimation of the source properties, are shown in the bottom panel of Fig. 1. The Virgo data are nonstationary above 150 Hz due to scattered light from the output optics modulated by alignment fluctuations and below 30 Hz due to seismic noise from anthropogenic activity. Occasional noise excess around the European power mains frequency of 50 Hz is also present. No noise subtraction was applied to the Virgo data prior to this analysis. The low signal amplitude observed in Virgo significantly constrained the sky position, but meant that the Virgo data did not contribute significantly to other parameters. As a result, the estimation of the source's parameters reported in Sec. IV is not impacted by the nonstationarity of Virgo data at the time of the event. Moreover, no unusual disturbance was observed by Virgo environmental sensors.

Data used in this study can be found in [80].

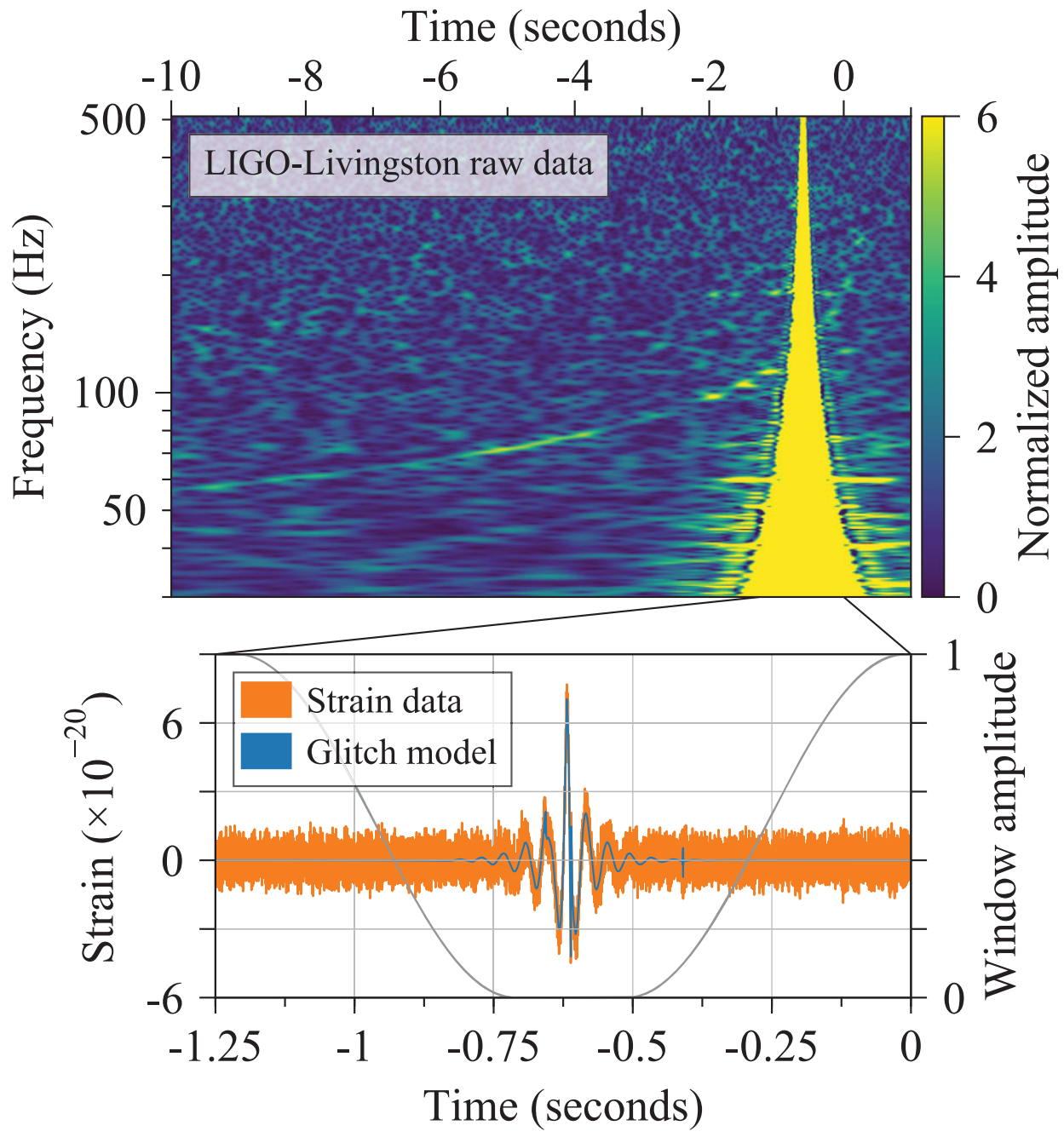

FIG. 2. Mitigation of the glitch in LIGO-Livingston data. Times are shown relative to August 17, 2017 12:41:04 UTC. *Top panel*: A time-frequency representation [65] of the raw LIGO-Livingston data used in the initial identification of GW170817 [76]. The coalescence time reported by the search is at time 0.4 s in this figure and the glitch occurs 1.1 s before this time. The time-frequency track of GW170817 is clearly visible despite the presence of the glitch. *Bottom panel*: The raw LIGO-Livingston strain data (orange curve) showing the glitch in the time domain. To mitigate the glitch in the rapid reanalysis that produced the sky map shown in Fig. 3 [77], the raw detector data were multiplied by an inverse Tukey window (gray curve, right axis) that zeroed out the data around the glitch [73]. To mitigate the glitch in the measurement of the source's properties, a model of the glitch based on a wavelet reconstruction [75] (blue curve) was subtracted from the data. The time-series data visualized in this figure have been bandpassed between 30 Hz and 2 kHz so that the detector's sensitive band is emphasized. The gravitational-wave strain amplitude of GW170817 is of the order of $10^{-22}$ and so is not visible in the bottom panel.

## III. DETECTION

GW170817 was initially identified as a single-detector event with the LIGO-Hanford detector by a low-latency binary-coalescence search [81–83] using template waveforms computed in post-Newtonian theory [11,13,36,84]. The two LIGO detectors and the Virgo detector were all taking data at the time; however, the saturation at the LIGO-Livingston detector prevented the search from registering a simultaneous event in both LIGO detectors, and the low-latency transfer of Virgo data was delayed.

Visual inspection of the LIGO-Hanford and LIGO-Livingston detector data showed the presence of a clear, long-duration chirp signal in time-frequency representations of the detector strain data. As a result, an initial alert was generated reporting a highly significant detection of a binary neutron star signal [85] in coincidence with the independently observed $\gamma$-ray burst GRB 170817A [39–41].

A rapid binary-coalescence reanalysis [86,87], with the time series around the glitch suppressed with a window function [73], as shown in Fig. 2, confirmed the presence of a significant coincident signal in the LIGO detectors. The source was rapidly localized to a region of 31 deg$^2$, shown in Fig. 3, using data from all three detectors [88]. This sky map was issued to observing partners, allowing the identification of an electromagnetic counterpart [46,48,50,77].

The combined SNR of GW170817 is estimated to be 32.4, with values 18.8, 26.4, and 2.0 in the LIGO-Hanford,

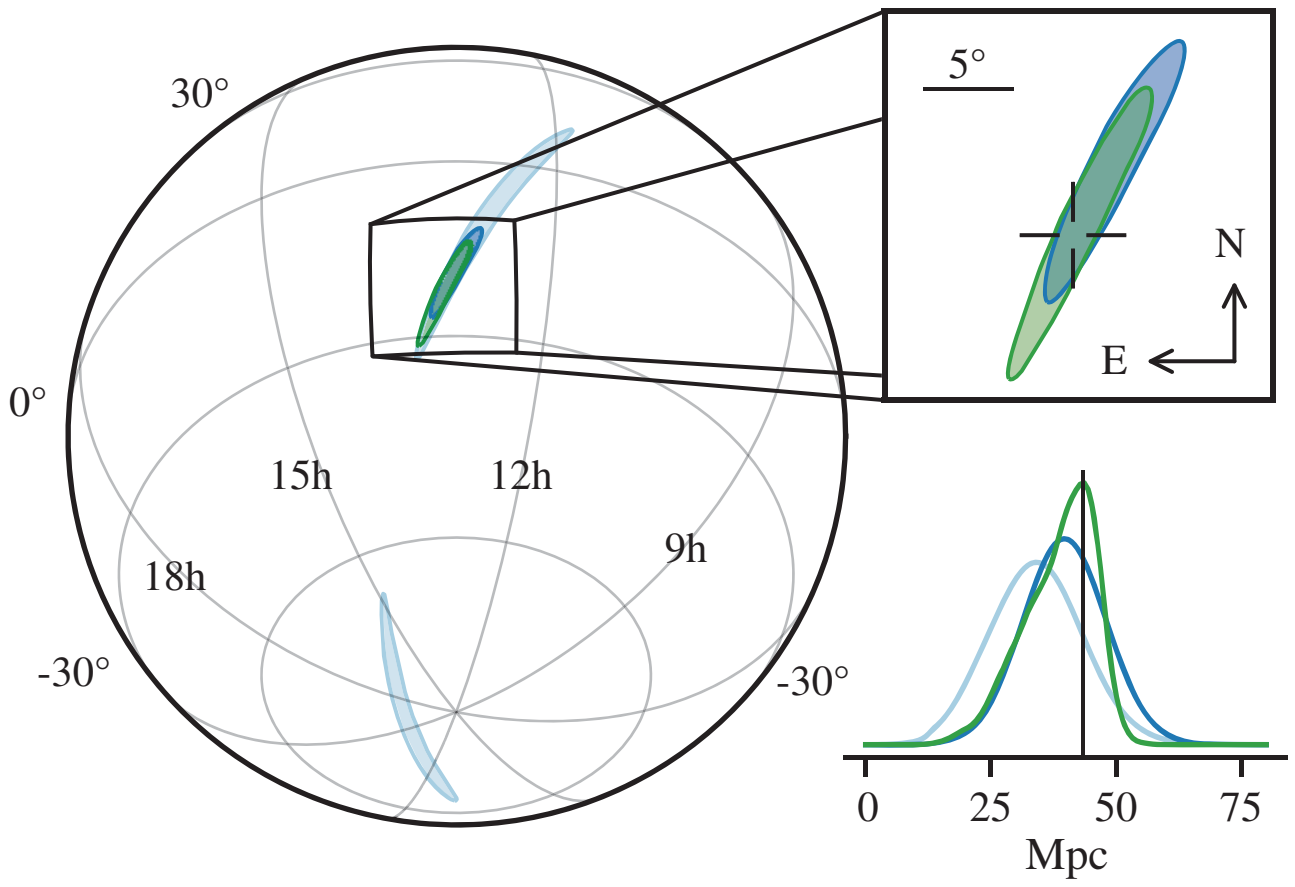

FIG. 3. Sky location reconstructed for GW170817 by a rapid localization algorithm from a Hanford-Livingston (190 deg$^2$, light blue contours) and Hanford-Livingston-Virgo (31 deg$^2$, dark blue contours) analysis. A higher latency Hanford-Livingston-Virgo analysis improved the localization (28 deg$^2$, green contours). In the top-right inset panel, the reticle marks the position of the apparent host galaxy NGC 4993. The bottom-right panel shows the *a posteriori* luminosity distance distribution from the three gravitational-wave localization analyses. The distance of NGC 4993, assuming the redshift from the NASA/IPAC Extragalactic Database [89] and standard cosmological parameters [90], is shown with a vertical line.

LIGO-Livingston, and Virgo data respectively, making it the loudest gravitational-wave signal so far detected. Two matched-filter binary-coalescence searches targeting sources with total mass between 2 and 500 $M_\odot$ in the detector frame were used to estimate the significance of this event [9,12,30,32,73,81–83,86,87,91–97]. The searches analyzed 5.9 days of LIGO data between August 13, 2017 02:00 UTC and August 21, 2017 01:05 UTC. Events are assigned a detection-statistic value that ranks their probability of being a gravitational-wave signal. Each search uses a different method to compute this statistic and measure the search background—the rate at which detector noise produces events with a detection-statistic value equal to or higher than the candidate event.

GW170817 was identified as the most significant event in the 5.9 days of data, with an estimated false alarm rate of one in $1.1 \times 10^6$ years with one search [81,83], and a consistent bound of less than one in $8.0 \times 10^4$ years for the other [73,86,87]. The second most significant signal in this analysis of 5.9 days of data is GW170814, which has a combined SNR of 18.3 [29]. Virgo data were not used in these significance estimates, but were used in the sky localization of the source and inference of the source properties.

## IV. SOURCE PROPERTIES

General relativity makes detailed predictions for the inspiral and coalescence of two compact objects, which may be neutron stars or black holes. At early times, for low orbital and gravitational-wave frequencies, the chirplike time evolution of the frequency is determined primarily by a specific combination of the component masses $m_1$ and $m_2$, the chirp mass $\mathcal{M} = (m_1 m_2)^{3/5}(m_1 + m_2)^{-1/5}$. As the orbit shrinks and the gravitational-wave frequency grows rapidly, the gravitational-wave phase is increasingly influenced by relativistic effects related to the mass ratio $q = m_2/m_1$, where $m_1 \geq m_2$, as well as spin-orbit and spin-spin couplings [98].

The details of the objects' internal structure become important as the orbital separation approaches the size of the bodies. For neutron stars, the tidal field of the companion induces a mass-quadrupole moment [99,100] and accelerates the coalescence [101]. The ratio of the induced quadrupole moment to the external tidal field is proportional to the tidal deformability (or polarizability) $\Lambda = (2/3)k_2[(c^2/G)(R/m)]^5$, where $k_2$ is the second Love number and $R$ is the stellar radius. Both $R$ and $k_2$ are fixed for a given stellar mass $m$ by the equation of state (EOS) for neutron-star matter, with $k_2 \simeq 0.05$–$0.15$ for realistic neutron stars [102–104]. Black holes are expected to have $k_2 = 0$ [99,105–109], so this effect would be absent.

As the gravitational-wave frequency increases, tidal effects in binary neutron stars increasingly affect the phase and become significant above $f_{\mathrm{GW}} \simeq 600$ Hz, so they are potentially observable [103,110–116]. Tidal deformabilities correlate with masses and spins, and our measurements are sensitive to the accuracy with which we describe the point-mass, spin, and tidal dynamics [113,117–119]. The point-mass dynamics has been calculated within the post-Newtonian framework [34,36,37], effective-one-body formalism [10,120–125], and with a phenomenological approach [126–131]. Results presented here are obtained using a frequency domain post-Newtonian waveform model [30] that includes dynamical effects from tidal interactions [132], point-mass spin-spin interactions [34,37,133,134], and couplings between the orbital angular momentum and the orbit-aligned dimensionless spin components of the stars $\chi_z$ [92].

The properties of gravitational-wave sources are inferred by matching the data with predicted waveforms. We perform a Bayesian analysis in the frequency range 30–2048 Hz that includes the effects of the $1\sigma$ calibration uncertainties on the received signal [135,136] ($< 7\%$ in amplitude and 3° in phase for the LIGO detectors [137] and 10% and 10° for Virgo at the time of the event). Unless otherwise specified, bounds on the properties of GW170817 presented in the text and in Table I are 90% posterior probability intervals that enclose systematic differences from currently available waveform models.

To ensure that the applied glitch mitigation procedure previously discussed in Sec. II (see Fig. 2) did not bias the estimated parameters, we added simulated signals with known parameters to data that contained glitches analogous to the one observed at the LIGO-Livingston detector during GW170817. After applying the glitch subtraction technique, we found that the bias in recovered parameters relative to their known values was well within their uncertainties. This can be understood by noting that a small time cut out of the coherent integration of the phase evolution has little impact on the recovered parameters. To corroborate these results, the test was also repeated with a window function applied, as shown in Fig. 2 [73].

TABLE I. Source properties for GW170817: we give ranges encompassing the 90% credible intervals for different assumptions of the waveform model to bound systematic uncertainty. The mass values are quoted in the frame of the source, accounting for uncertainty in the source redshift.

| | Low-spin priors ($\lvert\chi\rvert \le 0.05$) | High-spin priors ($\lvert\chi\rvert \le 0.89$) |
|---|---|---|
| Primary mass $m_1$ | 1.36–1.60 $M_\odot$ | 1.36–2.26 $M_\odot$ |
| Secondary mass $m_2$ | 1.17–1.36 $M_\odot$ | 0.86–1.36 $M_\odot$ |
| Chirp mass $\mathcal{M}$ | $1.188^{+0.004}_{-0.002} M_\odot$ | $1.188^{+0.004}_{-0.002} M_\odot$ |
| Mass ratio $m_2/m_1$ | 0.7–1.0 | 0.4–1.0 |
| Total mass $m_{\rm tot}$ | $2.74^{+0.04}_{-0.01} M_\odot$ | $2.82^{+0.47}_{-0.09} M_\odot$ |
| Radiated energy $E_{\rm rad}$ | $> 0.025 M_\odot c^2$ | $> 0.025 M_\odot c^2$ |
| Luminosity distance $D_{\rm L}$ | $40^{+8}_{-14}$ Mpc | $40^{+8}_{-14}$ Mpc |
| Viewing angle $\Theta$ | $\le 55°$ | $\le 56°$ |
| Using NGC 4993 location | $\le 28°$ | $\le 28°$ |
| Combined dimensionless tidal deformability $\tilde{\Lambda}$ | $\le 800$ | $\le 700$ |
| Dimensionless tidal deformability $\Lambda(1.4 M_\odot)$ | $\le 800$ | $\le 1400$ |

The source was localized to a region of the sky 28 $\mathrm{deg}^2$ in area, and 380 $\mathrm{Mpc}^3$ in volume, near the southern end of the constellation Hydra, by using a combination of the timing, phase, and amplitude of the source as observed in the three detectors [138,139]. The third detector, Virgo, was essential in localizing the source to a single region of the sky, as shown in Fig. 3. The small sky area triggered a successful follow-up campaign that identified an electromagnetic counterpart [50].

The luminosity distance to the source is $40^{+8}_{-14}$ Mpc, the closest ever observed gravitational-wave source and, by association, the closest short $\gamma$-ray burst with a distance measurement [45]. The distance measurement is correlated with the inclination angle $\cos\theta_{JN} = \hat{\mathbf{J}} \cdot \hat{\mathbf{N}}$, where $\hat{\mathbf{J}}$ is the unit vector in the direction of the total angular momentum of the system and $\hat{\mathbf{N}}$ is that from the source towards the observer [140]. We find that the data are consistent with an antialigned source: $\cos\theta_{JN} \le -0.54$, and the viewing angle $\Theta \equiv \min(\theta_{JN}, 180° - \theta_{JN})$ is $\Theta \le 56°$. Since the luminosity distance of this source can be determined independently of the gravitational wave data alone, we can use the association with NGC 4993 to break the distance degeneracy with $\cos\theta_{JN}$. The estimated Hubble flow velocity near NGC 4993 of $3017 \pm 166\ \mathrm{km\,s^{-1}}$ [141] provides a redshift, which in a flat cosmology with $H_0 = 67.90 \pm 0.55\ \mathrm{km\,s^{-1}\,Mpc^{-1}}$ [90], constrains $\cos\theta_{JN} < -0.88$ and $\Theta < 28°$. The constraint varies with the assumptions made about $H_0$ [141].

From the gravitational-wave phase and the ~3000 cycles in the frequency range considered, we constrain the chirp mass in the detector frame to be $\mathcal{M}^{\rm det} = 1.1977^{+0.0008}_{-0.0003} M_\odot$ [51]. The mass parameters in the detector frame are related to the rest-frame masses of the source by its redshift $z$ as $m^{\rm det} = m(1+z)$ [142]. Assuming the above cosmology [90], and correcting for the motion of the Solar System Barycenter with respect to the Cosmic Microwave Background [143], the gravitational-wave distance measurement alone implies a cosmological redshift of $0.008^{+0.002}_{-0.003}$, which is consistent with that of NGC 4993 [50,141,144,145]. Without the host galaxy, the uncertainty in the source's chirp mass $\mathcal{M}$ is dominated by the uncertainty in its luminosity distance. Independent of the waveform model or the choice of priors, described below, the source-frame chirp mass is $\mathcal{M} = 1.188^{+0.004}_{-0.002} M_\odot$.

While the chirp mass is well constrained, our estimates of the component masses are affected by the degeneracy between mass ratio $q$ and the aligned spin components $\chi_{1z}$ and $\chi_{2z}$ [38,146–150]. Therefore, the estimates of $q$ and the component masses depend on assumptions made about the admissible values of the spins. While $\chi < 1$ for black holes, and quark stars allow even larger spin values, realistic NS equations of state typically imply more stringent limits. For the set of EOS studied in [151] $\chi < 0.7$, although other EOS can exceed this bound. We began by assuming $\lvert\chi\rvert \le 0.89$, a limit imposed by available rapid waveform models, with an isotropic prior on the spin direction. With these priors we recover $q \in (0.4, 1.0)$ and a constraint on the effective aligned spin of the system [127,152] of $\chi_{\rm eff} \in (-0.01, 0.17)$. The aligned spin components are consistent with zero, with stricter bounds than in previous BBH observations [26,28,29]. Analysis using the effective precessing phenomenological waveforms of [128], which do not contain tidal effects, demonstrates that spin components in the orbital plane are not constrained.

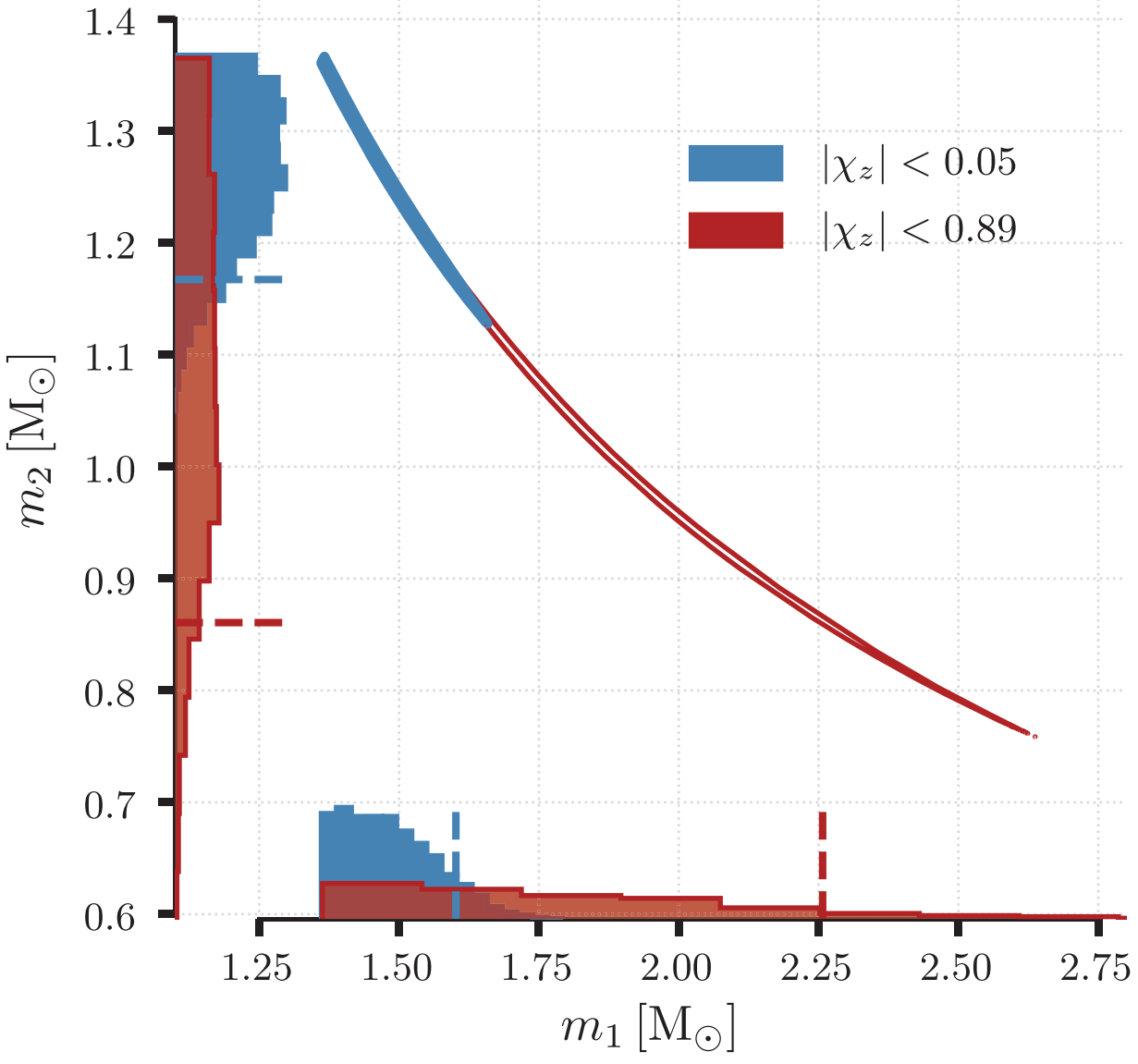

FIG. 4. Two-dimensional posterior distribution for the component masses $m_1$ and $m_2$ in the rest frame of the source for the low-spin scenario ($|\chi| < 0.05$, blue) and the high-spin scenario ($|\chi| < 0.89$, red). The colored contours enclose 90% of the probability from the joint posterior probability density function for $m_1$ and $m_2$. The shape of the two dimensional posterior is determined by a line of constant $\mathcal{M}$ and its width is determined by the uncertainty in $\mathcal{M}$. The widths of the marginal distributions (shown on axes, dashed lines enclose 90% probability away from equal mass of $1.36M_\odot$) is strongly affected by the choice of spin priors. The result using the low-spin prior (blue) is consistent with the masses of all known binary neutron star systems.

From $\mathcal{M}$ and $q$, we obtain a measure of the component masses $m_1 \in (1.36, 2.26)M_\odot$ and $m_2 \in (0.86, 1.36)M_\odot$, shown in Fig. 4. As discussed in Sec. I, these values are within the range of known neutron-star masses and below those of known black holes. In combination with electromagnetic observations, we regard this as evidence of the BNS nature of GW170817.

The fastest-spinning known neutron star has a dimensionless spin $\lesssim 0.4$ [153], and the possible BNS J1807-2500B has spin $\lesssim 0.2$ [154], after allowing for a broad range of equations of state. However, among BNS that will merge within a Hubble time, PSR J0737-3039A [155] has the most extreme spin, less than $\sim 0.04$ after spin-down is extrapolated to merger. If we restrict the spin magnitude in our analysis to $|\chi| \leq 0.05$, consistent with the observed population, we recover the mass ratio $q \in (0.7, 1.0)$ and component masses $m_1 \in (1.36, 1.60)M_\odot$ and $m_2 \in (1.17, 1.36)M_\odot$ (see Fig. 4). We also recover $\chi_{\mathrm{eff}} \in (-0.01, 0.02)$, where the upper limit is consistent with the low-spin prior.

Our first analysis allows the tidal deformabilities of the high-mass and low-mass component, $\Lambda_1$ and $\Lambda_2$, to vary independently. Figure 5 shows the resulting 90% and 50% contours on the posterior distribution with the post-Newtonian waveform model for the high-spin and low-spin priors. As a comparison, we show predictions coming from a set of candidate equations of state for neutron-star matter [156–160], generated using fits from [161]. All EOS support masses of $2.01 \pm 0.04 M_\odot$. Assuming that both components are neutron stars described by the same equation of state, a single function $\Lambda(m)$ is computed from the static $\ell = 2$ perturbation of a Tolman-Oppenheimer-Volkoff solution [103]. The shaded regions in Fig. 5 represent the values of the tidal deformabilities $\Lambda_1$ and $\Lambda_2$ generated using an equation of state from the 90% most probable fraction of the values of $m_1$ and $m_2$, consistent with the posterior shown in Fig. 4. We find that our constraints on $\Lambda_1$ and $\Lambda_2$ disfavor equations of state that predict less compact stars, since the mass range we recover generates $\Lambda$ values outside the 90% probability region. This is consistent with radius constraints from x-ray observations of neutron stars [162–166]. Analysis methods, in development, that *a priori* assume the same EOS governs both stars should improve our constraints [167].

To leading order in $\Lambda_1$ and $\Lambda_2$, the gravitational-wave phase is determined by the parameter

$$\tilde{\Lambda} = \frac{16}{13} \frac{(m_1 + 12m_2)m_1^4\Lambda_1 + (m_2 + 12m_1)m_2^4\Lambda_2}{(m_1 + m_2)^5} \quad (1)$$

[101,117]. Assuming a uniform prior on $\tilde{\Lambda}$, we place a 90% upper limit of $\tilde{\Lambda} \leq 800$ in the low-spin case and $\tilde{\Lambda} \leq 700$ in the high-spin case. We can also constrain the function $\Lambda(m)$ more directly by expanding $\Lambda(m)$ linearly about $m = 1.4M_\odot$ (as in [112,115]), which gives $\Lambda(1.4M_\odot) \leq 1400$ for the high-spin prior and $\Lambda(1.4M_\odot) \leq 800$ for the low-spin prior. A 95% upper bound inferred with the low-spin prior, $\Lambda(1.4M_\odot) \leq 970$, begins to compete with the 95% upper bound of 1000 derived from x-ray observations in [168].

Since the energy emitted in gravitational waves depends critically on the EOS of neutron-star matter, with a wide range consistent with constraints above, we are only able to place a lower bound on the energy emitted before the onset of strong tidal effects at $f_{\mathrm{GW}} \sim 600$ Hz as $E_{\mathrm{rad}} > 0.025 M_\odot c^2$. This is consistent with $E_{\mathrm{rad}}$ obtained from numerical simulations and fits for BNS systems consistent with GW170817 [114,169–171].

We estimate systematic errors from waveform modeling by comparing the post-Newtonian results with parameters recovered using an effective-one-body model [124] augmented with tidal effects extracted from numerical relativity with hydrodynamics [172]. This does not change the 90% credible intervals for component masses and effective spin under low-spin priors, but in the case of high-spin priors, we obtain the more restrictive $m_1 \in (1.36, 1.93)M_\odot$, $m_2 \in (0.99, 1.36)M_\odot$, and $\chi_{\mathrm{eff}} \in (0.0, 0.09)$. Recovered tidal deformabilities indicate shifts in the posterior distributions towards smaller values, with upper bounds for $\tilde{\Lambda}$ and $\Lambda(1.4M_\odot)$ reduced by a factor of roughly (0.8, 0.8) in the low-spin case and (1.0, 0.7) in the high-spin case. Further analysis is required to establish the uncertainties of these tighter bounds, and a detailed study of systematics is a subject of ongoing work.

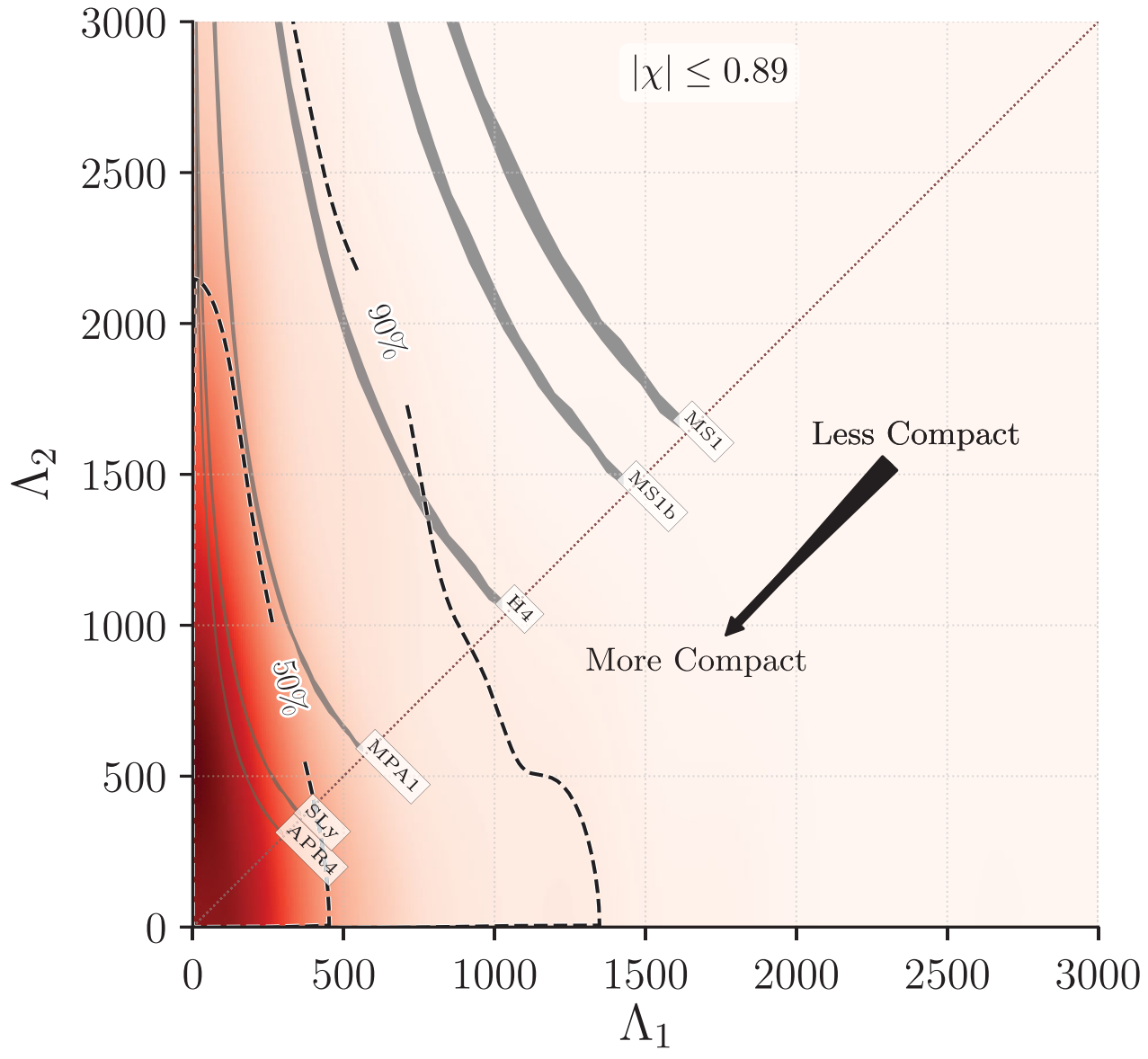

FIG. 5. Probability density for the tidal deformability parameters of the high and low mass components inferred from the detected signals using the post-Newtonian model. Contours enclosing 90% and 50% of the probability density are overlaid (dashed lines). The diagonal dashed line indicates the $\Lambda_1 = \Lambda_2$ boundary. The $\Lambda_1$ and $\Lambda_2$ parameters characterize the size of the tidally induced mass deformations of each star and are proportional to $k_2(R/m)^5$. Constraints are shown for the high-spin scenario $|\chi| \leq 0.89$ (left panel) and for the low-spin $|\chi| \leq 0.05$ (right panel). As a comparison, we plot predictions for tidal deformability given by a set of representative equations of state [156–160] (shaded filled regions), with labels following [161], all of which support stars of $2.01 M_\odot$. Under the assumption that both components are neutron stars, we apply the function $\Lambda(m)$ prescribed by that equation of state to the 90% most probable region of the component mass posterior distributions shown in Fig. 4. EOS that produce less compact stars, such as MS1 and MS1b, predict $\Lambda$ values outside our 90% contour.

Preliminary comparisons with waveform models under development [171,173–177] also suggest the post-Newtonian model used will systematically overestimate the value of the tidal deformabilities. Therefore, based on our current understanding of the physics of neutron stars, we consider the post-Newtonian results presented in this Letter to be conservative upper limits on tidal deformability. Refinements should be possible as our knowledge and models improve.

## V. IMPLICATIONS

### A. Astrophysical rate

Our analyses identified GW170817 as the only BNS-mass signal detected in O2 with a false alarm rate below 1/100 yr. Using a method derived from [27,178,179], and assuming that the mass distribution of the components of BNS systems is flat between 1 and 2 $M_\odot$ and their dimensionless spins are below 0.4, we are able to infer the local coalescence rate density $R$ of BNS systems. Incorporating the upper limit of 12600 $\mathrm{Gpc^{-3}\,yr^{-1}}$ from O1 as a prior, $R = 1540^{+3200}_{-1220}\ \mathrm{Gpc^{-3}\,yr^{-1}}$. Our findings are consistent with the rate inferred from observations of galactic BNS systems [19,20,155,180].

From this inferred rate, the stochastic background of gravitational wave s produced by unresolved BNS mergers throughout the history of the Universe should be comparable in magnitude to the stochastic background produced by BBH mergers [181,182]. As the advanced detector network improves in sensitivity in the coming years, the total stochastic background from BNS and BBH mergers should be detectable [183].

### B. Remnant

Binary neutron star mergers may result in a short- or long-lived neutron star remnant that could emit gravitational waves following the merger [184–190]. The ringdown of a black hole formed after the coalescence could also produce gravitational waves, at frequencies around 6 kHz, but the reduced interferometer response at high frequencies makes their observation unfeasible. Consequently, searches have been made for short (tens of ms) and intermediate duration ($\leq 500$ s) gravitational-wave signals from a neutron star remnant at frequencies up to 4 kHz [75,191,192]. For the latter, the data examined start at the time of the coalescence and extend to the end of the observing run on August 25, 2017. With the time scales and methods considered so far [193], there is no evidence of a postmerger signal of astrophysical origin. However, upper limits placed on the strength of gravitational-wave emission cannot definitively rule out the existence of a short- or long-lived postmerger neutron star. The implications of various postmerger scenarios are explored in [45,193].

### C. Tests of gravity

GRB 170817A was observed 1.7 s after GW170817. Combining this delay with the knowledge of the source luminosity distance, strong constraints are placed on the fundamental physics of gravity. The observed arrival times are used to investigate the speed of gravity, Lorentz invariance, and tests of the equivalence principle through the Shapiro time delay, as reported in [45].

We also expect the much longer duration of the BNS signal compared to previous BBH gravitational-wave sources to yield significantly improved constraints when testing for waveform deviations from general relativity using a parametrized waveform expansion [194], especially at low post-Newtonian orders. Placing these bounds requires a deep understanding of the systematic uncertainties resulting from waveform modeling and data conditioning, and is the subject of ongoing investigations.

### D. Cosmology

The gravitational-wave signal gives a direct measurement of the luminosity distance of the source, which, along with a redshift measurement, can be used to infer cosmological parameters independently of the cosmic distance ladder [141,195]. Using the association with the galaxy NGC 4993 and the luminosity distance directly measured from the gravitational-wave signal, the Hubble constant is inferred to be $H_0 = 70^{+12}_{-8}\ \mathrm{km\,s^{-1}\,Mpc^{-1}}$ [141] (most probable value and minimum 68.3% probability range, which can be compared to the value from Planck $H_0 = 67.90 \pm 0.55\ \mathrm{km\,s^{-1}\,Mpc^{-1}}$ [90]). Alternatively, we may assume the cosmology is known and use the association with NGC 4993 to constrain the luminosity distance of the source, in which case the gravitational-wave measurement of the inclination angle of the source is significantly improved, with consequences for the $\gamma$-ray burst opening angle and related physics [45].

## VI. CONCLUSIONS

In this Letter we have presented the first detection of gravitational waves from the inspiral of a binary neutron star system. Gravitational-wave event GW170817, observed and localized by the two Advanced LIGO detectors and the Advanced Virgo detector, is the loudest gravitational-wave signal detected to date. This coalescence event was followed by a short burst of $\gamma$ rays observed with the *Fermi* Gamma-Ray Burst Monitor [39–42] and *INTEGRAL* [43,44]. The coincident observation of a gravitational-wave signal and a $\gamma$-ray burst appears to confirm the long-held hypothesis that BNS mergers are linked to short-$\gamma$-ray bursts [196,197]. Subsequent observations have determined the location of the source and followed its evolution through the electromagnetic spectrum [50].

Detailed analyses of the gravitational-wave data, together with observations of electromagnetic emissions, are providing new insights into the astrophysics of compact binary systems and $\gamma$-ray bursts, dense matter under extreme conditions, the nature of gravitation, and independent tests of cosmology. Less than two years after the debut of gravitational-wave astronomy, GW170817 marks the beginning of a new era of discovery.

## ACKNOWLEDGMENTS

The authors gratefully acknowledge the support of the United States National Science Foundation (NSF) for the construction and operation of the LIGO Laboratory and Advanced LIGO as well as the Science and Technology Facilities Council (STFC) of the United Kingdom, the Max-Planck-Society (MPS), and the State of Niedersachsen, Germany for support of the construction of Advanced LIGO and construction and operation of the GEO600 detector. Additional support for Advanced LIGO was provided by the Australian Research Council. The authors gratefully acknowledge the Italian Istituto Nazionale di Fisica Nucleare (INFN), the French Centre National de la Recherche Scientifique (CNRS), and the Foundation for Fundamental Research on Matter supported by the Netherlands Organisation for Scientific Research, for the construction and operation of the Virgo detector and the creation and support of the EGO consortium. The authors also gratefully acknowledge research support from these agencies as well as by the Council of Scientific and Industrial Research of India, the Department of Science and Technology, India, the Science & Engineering Research Board (SERB), India, the Ministry of Human Resource Development, India, the Spanish Agencia Estatal de Investigación, the Vicepresidència i Conselleria d'Innovació, Recerca i Turisme and the Conselleria d'Educació i Universitat del Govern de les Illes Balears, the Conselleria d'Educació, Investigació, Cultura i Esport de la Generalitat Valenciana, the National Science Centre of Poland, the Swiss National Science Foundation (SNSF), the Russian Foundation for Basic Research, the Russian Science Foundation, the European Commission, the European Regional Development Funds (ERDF), the Royal Society, the Scottish Funding Council, the Scottish Universities Physics Alliance, the Hungarian Scientific Research Fund (OTKA), the Lyon Institute of Origins (LIO), the National Research, Development and Innovation Office Hungary (NKFI), the National Research Foundation of Korea, Industry Canada and the Province of Ontario through the Ministry of Economic Development and Innovation, the Natural Science and Engineering Research Council Canada, the Canadian Institute for Advanced Research, the Brazilian

Ministry of Science, Technology, Innovations, and Communications, the International Center for Theoretical Physics South American Institute for Fundamental Research (ICTP-SAIFR), the Research Grants Council of Hong Kong, the National Natural Science Foundation of China (NSFC), the Leverhulme Trust, the Research Corporation, the Ministry of Science and Technology (MOST), Taiwan and the Kavli Foundation. The authors gratefully acknowledge the support of the NSF, STFC, MPS, INFN, CNRS and the State of Niedersachsen, Germany for provision of computational resources. This research has made use of the NASA/IPAC Extragalactic Database (NED) which is operated by the Jet Propulsion Laboratory, California Institute of Technology, under contract with the National Aeronautics and Space Administration.

[1] R. A. Hulse and J. H. Taylor, Astrophys. J. **195**, L51 (1975).
[2] J. H. Taylor and J. M. Weisberg, Astrophys. J. **253**, 908 (1982).
[3] R. Weiss, MIT Technical Report No. LIGO-P720002, 1972.
[4] R. W. P. Drever, in *Gravitational Radiation*, edited by N. Deruelle and T. Piran (North-Holland, Amsterdam, 1983), p. 321.
[5] A. Brillet *et al.* (Virgo Collaboration), Technical Report No. VIR-0517A-15, 1989 [https://tds.virgo-gw.eu/?content=3&r=12122].
[6] J. Hough *et al.*, MPQ Technical Report No. 147, 1989 [GWD/137/JH(89)].
[7] K. Thorne, *Three Hundred Years of Gravitation*, edited by S. W. Hawking and W. Israel (Cambridge University Press, Cambridge, U.K., 1987), pp. 330–458.
[8] R. N. Manchester, G. B. Hobbs, A. Teoh, and M. Hobbs, Astron. J. **129**, 1993 (2005).
[9] L. Blanchet, Thibault Damour, Bala R. Iyer, Clifford M. Will, and Alan G. Wiseman, Phys. Rev. Lett. **74**, 3515 (1995).
[10] A. Buonanno and T. Damour, Phys. Rev. D **59**, 084006 (1999).
[11] T. Damour, P. Jaranowski, and G. Schäfer, Phys. Lett. B **513**, 147 (2001).
[12] L. Blanchet, T. Damour, G. Esposito-Farese, and B. R. Iyer, Phys. Rev. Lett. **93**, 091101 (2004).
[13] W. D. Goldberger and I. Z. Rothstein, Phys. Rev. D **73**, 104029 (2006).
[14] B. P. Flannery and E. P. J. van den Heuvel, Astron. Astrophys. **39**, 61 (1975).
[15] A. G. Massevitch, A. V. Tutukov, and L. R. Yungelson, Astrophys. Space Sci. **40**, 115 (1976).
[16] J. P. A. Clark, E. P. J. van den Heuvel, and W. Sutantyo, Astron. Astrophys. **72**, 120 (1979).
[17] E. S. Phinney, Astrophys. J. Lett. **380**, L17 (1991).
[18] V. Kalogera *et al.*, Astrophys. J. Lett. **601**, L179 (2004). V. Kalogera *et al.*, Astrophys. J. Lett. **614**, L137(E) (2004).
[19] J. Abadie *et al.* (LIGO Scientific Collaboration and Virgo Collaboration), Classical Quantum Gravity **27**, 173001 (2010).
[20] C. Kim, B. B. P. Perera, and M. A. McLaughlin, Mon. Not. R. Astron. Soc. **448**, 928 (2015).
[21] B. P. Abbott *et al.* (KAGRA Collaboration, LIGO Scientific Collaboration, Virgo Collaboration), Living Rev. Relativity **19**, 1 (2016).
[22] J. Aasi *et al.* (LIGO Scientific Collaboration), Classical Quantum Gravity **32**, 074001 (2015).
[23] F. Acernese *et al.* (Virgo Collaboration), Classical Quantum Gravity **32**, 024001 (2015).
[24] B. P. Abbott *et al.* (LIGO Scientific Collaboration and Virgo Collaboration), Phys. Rev. Lett. **116**, 061102 (2016).
[25] B. P. Abbott *et al.* (LIGO Scientific Collaboration and Virgo Collaboration), Phys. Rev. Lett. **116**, 241103 (2016).
[26] B. P. Abbott *et al.* (LIGO Scientific Collaboration and Virgo Collaboration), Phys. Rev. X **6**, 041015 (2016).
[27] B. P. Abbott *et al.* (LIGO Scientific Collaboration and Virgo Collaboration), Astrophys. J. Lett. **832**, L21 (2016).
[28] B. P. Abbott *et al.* (LIGO Scientific Collaboration and Virgo Collaboration), Phys. Rev. Lett. **118**, 221101 (2017).
[29] B. P. Abbott *et al.* (LIGO Scientific Collaboration and Virgo Collaboration), Phys. Rev. Lett. **119**, 141101 (2017).
[30] B. S. Sathyaprakash and S. V. Dhurandhar, Phys. Rev. D **44**, 3819 (1991).
[31] C. Cutler *et al.*, Phys. Rev. Lett. **70**, 2984 (1993).
[32] B. Allen, W. G. Anderson, P. R. Brady, D. A. Brown, and J. D. E. Creighton, Phys. Rev. D **85**, 122006 (2012).
[33] D. A. Brown, P. Kumar, and A. H. Nitz, Phys. Rev. D **87**, 082004 (2013).
[34] K. G. Arun, A. Buonanno, G. Faye, and E. Ochsner, Phys. Rev. D **79**, 104023 (2009); **84**, 049901(E) (2011).
[35] A. Buonanno, B. R. Iyer, E. Ochsner, Y. Pan, and B. S. Sathyaprakash, Phys. Rev. D **80**, 084043 (2009).
[36] L. Blanchet, Living Rev. Relativity **17**, 2 (2014).
[37] C. K. Mishra, A. Kela, K. G. Arun, and G. Faye, Phys. Rev. D **93**, 084054 (2016).
[38] C. Cutler and E. E. Flanagan, Phys. Rev. D **49**, 2658 (1994).
[39] A. G. A. von Kienlin, C. Meegan, and the Fermi GBM Team, GCN **21520**, 1 (2017).
[40] V. Connaughton *et al.*, GCN **21506**, 1 (2017).
[41] A. Goldstein *et al.*, GCN **21528**, 1 (2017).
[42] A. Goldstein *et al.*, Astrophys. J. Lett. **848**, L14 (2017).
[43] V. Savchenko *et al.*, GCN **21507**, 1 (2017).
[44] V. Savchenko *et al.*, Astrophys. J. Lett. **848**, L15 (2017).
[45] B. P. Abbott *et al.* (LIGO Scientific Collaboration and Virgo Collaboration), Astrophys. J. Lett. **848**, L13 (2017);
[46] D. A. Coulter *et al.*, GCN **21529**, 1 (2017).
[47] R. J. Foley *et al.*, GCN **21536**, 1 (2017).
[48] D. A. Coulter *et al.*, Science, in press (2017), DOI: 10.1126/science.aap9811.
[49] Y.-C. Pan *et al.*, Astrophys. J. Lett. **848**, L30 (2017).
[50] B. P. Abbott *et al.* (LIGO Scientific Collaboration and Virgo Collaboration), Astrophys. J. Lett. **848**, L12 (2017).
[51] P. C. Peters and J. Mathews, Phys. Rev. **131**, 435 (1963).
[52] T. M. Tauris *et al.*, Astrophys. J. **846**, 170 (2017).
[53] J. Antoniadis *et al.*, Science **340**, 1233232 (2013).

[54] F. Özel, D. Psaltis, R. Narayan, and J. E. McClintock, Astrophys. J. **725**, 1918 (2010).
[55] L. Kreidberg, C. D. Bailyn, W. M. Farr, and V. Kalogera, Astrophys. J. **757**, 36 (2012).
[56] R. Narayan and J. E. McClintock, in *General Relativity and Gravitation: A Centennial Perspective*, edited by A. Ashtekar, B. K. Berger, J. Isenberg, and M. MacCallum (Cambridge University Press, Cambridge, U.K., 2015), Chap. 3.3, pp. 133–147.
[57] D. J. Kaup, Phys. Rev. **172**, 1331 (1968).
[58] P. O. Mazur and E. Mottola, Proc. Natl. Acad. Sci. U.S.A. **101**, 9545 (2004).
[59] N. Itoh, Prog. Theor. Phys. **44**, 291 (1970).
[60] V. Cardoso, S. Hopper, C. F. B. Macedo, C. Palenzuela, and P. Pani, Phys. Rev. D **94**, 084031 (2016).
[61] V. Cardoso, E. Franzin, A. Maselli, P. Pani, and G. Raposo, Phys. Rev. D **95**, 084014 (2017); **95**, 089901 (2017).
[62] L. S. Finn and D. F. Chernoff, Phys. Rev. D **47**, 2198 (1993).
[63] H.-Y. Chen, D. E. Holz, J. Miller, M. Evans, S. Vitale, and J. Creighton, arXiv:1709.08079.
[64] K. L. Dooley *et al.*, Classical Quantum Gravity **33**, 075009 (2016).
[65] S. Chatterji, L Blackburn, G Martin, and E Katsavounidis, Classical Quantum Gravity **21**, S1809 (2004).
[66] R. G. Brown and P. Y. C. Hwang, *Introduction to Random Signals and Applied Kalman Filtering with Matlab Exercises* (Wiley, New York, 2012).
[67] J.C. Driggers *et al.*, Technical Report No. LIGO-P1700260, 2017.
[68] J. C. Driggers, M. Evans, K. Pepper, and R. Adhikari, Rev. Sci. Instrum. **83**, 024501 (2012).
[69] G.D. Meadors, K. Kawabe, and K. Riles, Classical Quantum Gravity **31**, 105014 (2014).
[70] V. Tiwari *et al.*, Classical Quantum Gravity **32**, 165014 (2015).
[71] L. Nuttall *et al.*, Classical Quantum Gravity **32**, 245005 (2015).
[72] B. P. Abbott *et al.* (LIGO Scientific Collaboration and Virgo Collaboration), Phys. Rev. D **93**, 122003 (2016).
[73] S. A. Usman *et al.*, Classical Quantum Gravity **33**, 215004 (2016).
[74] B. P. Abbott *et al.* (LIGO Scientific Collaboration and Virgo Collaboration), Technical Report No. LIGO-P1600110, 2017.
[75] N. J. Cornish and T. B. Littenberg, Classical Quantum Gravity **32**, 135012 (2015).
[76] LIGO Scientific Collaboration and Virgo Collaboration, GCN **21509**, 1 (2017).
[77] LIGO Scientific Collaboration and Virgo Collaboration, GCN **21513**, 1 (2017).
[78] B. P. Abbott *et al.* (LIGO Scientific Collaboration and Virgo Collaboration), Classical Quantum Gravity **33**, 134001 (2016).
[79] A. Effler, R M S Schofield, V V Frolov, G González, K Kawabe, J R Smith, J Birch, and R McCarthy, Classical Quantum Gravity **32**, 035017 (2015).
[80] LIGO Scientific Collaboration and Virgo Collaboration, LIGO Open Science Center (LOSC), 2017 [https://doi.org/10.7935/K5B8566F].
[81] K. Cannon *et al.*, Astrophys. J. **748**, 136 (2012).
[82] S. Privitera, S. R. P. Mohapatra, P. Ajith, K. Cannon, N. Fotopoulos, M. A. Frei, C. Hanna, A. J. Weinstein, and J. T. Whelan, Phys. Rev. D **89**, 024003 (2014).
[83] C. Messick *et al.*, Phys. Rev. D **95**, 042001 (2017).
[84] A. Einstein, L. Infeld, and B. Hoffmann, Ann. Math. **39**, 65 (1938).
[85] LIGO Scientific Collaboration and Virgo Collaboration, GCN **21505**, 1 (2017).
[86] A. H. Nitz, T. Dent, T. Dal Canton, S. Fairhurst, and D. A. Brown, arXiv:1705.01513.
[87] A. Nitz *et al.*, ligo-cbc/pycbc: O2 production release 20, 2017.
[88] L. P. Singer and L. R. Price, Phys. Rev. D **93**, 024013 (2016).
[89] NASA/IPAC Extragalactic Database.
[90] P. A. R. Ade *et al.* (Planck Collaboration), Astron. Astrophys. **594**, A13 (2016).
[91] L. Blanchet, A. Buonanno, and G. Faye, Phys. Rev. D **74**, 104034 (2006); **75**, 049903(E) (2007); **81**, 089901(E) (2010).
[92] A. Bohé, S. Marsat, and L. Blanchet, Classical Quantum Gravity **30**, 135009 (2013).
[93] B. J. Owen and B. S. Sathyaprakash, Phys. Rev. D **60**, 022002 (1999).
[94] D. A. Brown, I. Harry, A. Lundgren, and A. H. Nitz, Phys. Rev. D **86**, 084017 (2012).
[95] T. Dal Canton and I. W. Harry, arXiv:1705.01845.
[96] B. Allen, Phys. Rev. D **71**, 062001 (2005).
[97] A. H. Nitz, arXiv:1709.08974.
[98] C. Cutler, T. A. Apostolatos, L. Bildsten, L. S. Finn, E. E. Flanagan, D. Kennefick, D. M. Markovic, A. Ori, E. Poisson, G. J. Sussman, and K. S. Thorne, Phys. Rev. Lett. **70**, 2984 (1993).
[99] T. Damour, Lect. Notes Phys., **124**, 59 (1983).
[100] T. Damour, M. Soffel, and C.-m. Xu, Phys. Rev. D **45**, 1017 (1992).
[101] E. E. Flanagan and T. Hinderer, Phys. Rev. D **77**, 021502 (2008).
[102] T. Hinderer, Astrophys. J. **677**, 1216 (2008).
[103] T. Hinderer, B. D. Lackey, R. N. Lang, and J. S. Read, Phys. Rev. D **81**, 123016 (2010).
[104] S. Postnikov, M. Prakash, and J. M. Lattimer, Phys. Rev. D **82**, 024016 (2010).
[105] T. Damour and A. Nagar, Phys. Rev. D **80**, 084035 (2009).
[106] T. Binnington and E. Poisson, Phys. Rev. D **80**, 084018 (2009).
[107] B. Kol and M. Smolkin, J. High Energy Phys. 02 (2012) 010.
[108] P. Pani, L. Gualtieri, A. Maselli, and V. Ferrari, Phys. Rev. D **92**, 024010 (2015).
[109] P. Landry and E. Poisson, Phys. Rev. D **91**, 104018 (2015).
[110] J. S. Read, C. Markakis, M. Shibata, K. Uryu, J. D. E. Creighton, and J. L. Friedman, Phys. Rev. D **79**, 124033 (2009).
[111] T. Damour, A. Nagar, and L. Villain, Phys. Rev. D **85**, 123007 (2012).
[112] W. Del Pozzo, T. G. F. Li, M. Agathos, C. Van Den Broeck, and S. Vitale, Phys. Rev. Lett. **111**, 071101 (2013).

[113] L. Wade, J. D. E. Creighton, E. Ochsner, B. D. Lackey, B. F. Farr, T. B. Littenberg, and V. Raymond, Phys. Rev. D **89**, 103012 (2014).

[114] S. Bernuzzi, A. Nagar, S. Balmelli, T. Dietrich, and M. Ujevic, Phys. Rev. Lett. **112**, 201101 (2014).

[115] M. Agathos, J. Meidam, W. Del Pozzo, T. G. F. Li, M. Tompitak, J. Veitch, S. Vitale, and C. Van Den Broeck, Phys. Rev. D **92**, 023012 (2015).

[116] K. Hotokezaka, K. Kyutoku, Y.-i. Sekiguchi, and M. Shibata, Phys. Rev. D **93**, 064082 (2016).

[117] M. Favata, Phys. Rev. Lett. **112**, 101101 (2014).

[118] K. Yagi and N. Yunes, Phys. Rev. D **89**, 021303 (2014).

[119] T. Cullen, I. Harry, J. Read, and E. Flynn, arXiv: 1708.04359.

[120] A. Buonanno and T. Damour, Phys. Rev. D **62**, 064015 (2000).

[121] T. Damour, P. Jaranowski, and G. Schäfer, Phys. Rev. D **78**, 024009 (2008).

[122] T. Damour and A. Nagar, Phys. Rev. D **79**, 081503 (2009).

[123] E. Barausse and A. Buonanno, Phys. Rev. D **81**, 084024 (2010).

[124] A. Bohé *et al.*, Phys. Rev. D **95**, 044028 (2017).

[125] A. Nagar, G. Riemenschneider, and G. Pratten, arXiv:1703.06814 [Phys. Rev. D (to be published)].

[126] P. Ajith *et al.*, Classical Quantum Gravity **24**, S689 (2007).

[127] P. Ajith *et al.*, Phys. Rev. Lett. **106**, 241101 (2011).

[128] M. Hannam, Patricia Schmidt, Alejandro Bohé, Leïla Haegel, Sascha Husa, Frank Ohme, Geraint Pratten, and Michael Pürrer, Phys. Rev. Lett. **113**, 151101 (2014).

[129] S. Husa, S. Khan, M. Hannam, M. Purrer, F. Ohme, X. J. Forteza, and A. Bohe, Phys. Rev. D **93**, 044006 (2016).

[130] S. Khan, S. Husa, M. Hannam, F. Ohme, M. Purrer, X. J. Forteza, and A. Bohe, Phys. Rev. D **93**, 044007 (2016).

[131] R. Smith, S. E. Field, K. Blackburn, C.-J. Haster, M. Pürrer, V. Raymond, and P. Schmidt, Phys. Rev. D **94**, 044031 (2016).

[132] J. Vines, E. E. Flanagan, and T. Hinderer, Phys. Rev. D **83**, 084051 (2011).

[133] B. Mikoczi, M. Vasuth, and L. A. Gergely, Phys. Rev. D **71**, 124043 (2005).

[134] A. Bohé, G. Faye, S. Marsat, and E. K. Porter, Classical Quantum Gravity **32**, 195010 (2015).

[135] J. Veitch *et al.*, Phys. Rev. D **91**, 042003 (2015).

[136] B. P. Abbott *et al.* (LIGO Scientific Collaboration and Virgo Collaboration), Phys. Rev. Lett. **116**, 241102 (2016).

[137] C. Cahillane *et al.*, arXiv:1708.03023.

[138] L. P. Singer, H.-Y. Chen, D. E. Holz, W. M. Farr, L. R. Price, V. Raymond, S. B. Cenko, N. Gehrels, J. Cannizzo, M. M. Kasliwal, S. Nissanke, M. Coughlin, B. Farr, A. L. Urban, S. Vitale, J. Veitch, P. Graff, C. P. L. Berry, S. Mohapatra, and I. Mandel, Astrophys. J. Lett. **829**, L15 (2016).

[139] H.-Y. Chen and D. E. Holz, arXiv:1612.01471.

[140] T. A. Apostolatos, C. Cutler, G. J. Sussman, and K. S. Thorne, Phys. Rev. D **49**, 6274 (1994).

[141] B. P. Abbott *et al.* (LIGO Scientific Collaboration, Virgo Collaboration, IM2H Collaboration, Dark Energy Camera GW-EM Collaboration, DES Collaboration, DLT40 Collaboration, Las Cumbres Observatory Collaboration, VINROUGE Collaboration, and MASTER Collaboration), Nature (London), in press (2017), DOI: 10.1038/nature24471.

[142] A. Krolak and B. F. Schutz, Gen. Relativ. Gravit. **19**, 1163 (1987).

[143] D. J. Fixsen, Astrophys. J. **707**, 916 (2009).

[144] A. C. Crook, J. P. Huchra, N. Martimbeau, K. L. Masters, T. Jarrett, and L. M. Macri, Astrophys. J. **655**, 790 (2007).

[145] A. C. Crook, J. P. Huchra, N. Martimbeau, K. L. Masters, T. Jarrett, and L. M. Macri, Astrophys. J. **685**, 1320 (2008).

[146] M. Hannam, D. A. Brown, S. Fairhurst, C. L. Fryer, and I. W. Harry, Astrophys. J. Lett. **766**, L14 (2013).

[147] F. Ohme, A. B. Nielsen, D. Keppel, and A. Lundgren, Phys. Rev. D **88**, 042002 (2013).

[148] C. P. L. Berry *et al.*, Astrophys. J. **804**, 114 (2015).

[149] M. Pürrer, M. Hannam, and F. Ohme, Phys. Rev. D **93**, 084042 (2016).

[150] B. Farr *et al.*, Astrophys. J. **825**, 116 (2016).

[151] K.-W. Lo and L.-M. Lin, Astrophys. J. **728**, 12 (2011).

[152] E. Racine, Phys. Rev. D **78**, 044021 (2008).

[153] J. W. T. Hessels, S. M. Ransom, I. H. Stairs, P. C. C. Freire, V. M. Kaspi, and F. Camilo, Science **311**, 1901 (2006).

[154] R. S. Lynch, P. C. C. Freire, S. M. Ransom, and B. A. Jacoby, Astrophys. J. **745**, 109 (2012).

[155] M. Burgay *et al.*, Nature (London) **426**, 531 (2003).

[156] H. Müther, M. Prakash, and T. L. Ainsworth, Phys. Lett. B **199**, 469 (1987).

[157] H. Mueller and B. D. Serot, Nucl. Phys. A **A606**, 508 (1996).

[158] F. Douchin and P. Haensel, Astron. Astrophys. **380**, 151 (2001).

[159] A. Akmal, V. R. Pandharipande, and D. G. Ravenhall, Phys. Rev. C **58**, 1804 (1998).

[160] B. D. Lackey, M. Nayyar, and B. J. Owen, Phys. Rev. D **73**, 024021 (2006).

[161] J. S. Read, B. D. Lackey, B. J. Owen, and J. L. Friedman, Phys. Rev. D **79**, 124032 (2009).

[162] J. M. Lattimer and M. Prakash, Phys. Rep. **621**, 127 (2016).

[163] F. Ozel and P. Freire, Annu. Rev. Astron. Astrophys. **54**, 401 (2016).

[164] A. L. Watts *et al.*, Rev. Mod. Phys. **88**, 021001 (2016).

[165] A. W. Steiner, C. O. Heinke, S. Bogdanov, C. Li, W. C. G. Ho, A. Bahramian, and S. Han, arXiv:1709.05013.

[166] J. Nättilä, M. C. Miller, A. W. Steiner, J. J. E. Kajava, V. F. Suleimanov, and J. Poutanen, arXiv:1709.09120.

[167] B. D. Lackey and L. Wade, Phys. Rev. D **91**, 043002 (2015).

[168] A. W. Steiner, S. Gandolfi, F. J. Fattoyev, and W. G. Newton, Phys. Rev. C **91**, 015804 (2015).

[169] R. De Pietri, A. Feo, F. Maione, and F. Löffler, Phys. Rev. D **93**, 064047 (2016).

[170] T. Dietrich, M. Ujevic, W. Tichy, S. Bernuzzi, and B. Brügmann, Phys. Rev. D **95**, 024029 (2017).

[171] T. Dietrich and T. Hinderer, Phys. Rev. D **95**, 124006 (2017).

[172] T. Dietrich, S. Bernuzzi, and W. Tichy, arXiv:1706.02969.

[173] T. Damour and A. Nagar, Phys. Rev. D **81**, 084016 (2010).

[174] B. D. Lackey, S. Bernuzzi, C. R. Galley, J. Meidam, and C. Van Den Broeck, Phys. Rev. D **95**, 104036 (2017).

[175] S. Bernuzzi, A. Nagar, T. Dietrich, and T. Damour, Phys. Rev. Lett. **114**, 161103 (2015).
[176] T. Hinderer *et al.*, Phys. Rev. Lett. **116**, 181101 (2016).
[177] J. Steinhoff, T. Hinderer, A. Buonanno, and A. Taracchini, Phys. Rev. D **94**, 104028 (2016).
[178] B. P. Abbott *et al.* (LIGO Scientific Collaboration and Virgo Collaboration), Astrophys. J. Lett. **833**, L1 (2016).
[179] B. P. Abbott *et al.* (LIGO Scientific Collaboration and Virgo Collaboration), Astrophys. J. Suppl. Ser. **227**, 14 (2016).
[180] C. Kim, V. Kalogera, and D. R. Lorimer, Astrophys. J. **584**, 985 (2003).
[181] B. P. Abbott *et al.* (LIGO Scientific Collaboration and Virgo Collaboration), Phys. Rev. Lett. **118**, 121101 (2017).
[182] B. P. Abbott *et al.* (LIGO Scientific Collaboration and Virgo Collaboration), Phys. Rev. Lett. **116**, 131102 (2016).
[183] LIGO Scientific Collaboration and Virgo Collaboration, Technical Report No. LIGO-P170272, 2017.
[184] N. Andersson, Classical Quantum Gravity **20**, R105 (2003).
[185] A. L. Piro, B. Giacomazzo, and R. Perna, Astrophys. J. **844**, L19 (2017).
[186] J. A. Clark, A. Bauswein, N. Stergioulas, and D. Shoemaker, Classical Quantum Gravity **33**, 085003 (2016).
[187] P. D. Lasky and K. Glampedakis, Mon. Not. R. Astron. Soc. **458**, 1660 (2016).
[188] S. Bose, K. Chakravarti, L. Rezzolla, B. S. Sathyaprakash, and K. Takami, arXiv:1705.10850.
[189] M. Shibata and K. Uryū, Phys. Rev. D **61**, 064001 (2000).
[190] L. Baiotti and L. Rezzolla, Rep. Prog. Phys. **80**, 096901 (2017).
[191] S. Klimenko *et al.*, Phys. Rev. D **93**, 042004 (2016).
[192] B. P. Abbott *et al.* (LIGO Scientific Collaboration and Virgo Collaboration), Phys. Rev. D **93**, 042005 (2016).
[193] LIGO Scientific Collaboration and Virgo Collaboration *et al.*, Technical Report No. LIGO-P1700318, 2017.
[194] B. P. Abbott *et al.* (LIGO Scientific Collaboration and Virgo Collaboration), Phys. Rev. Lett. **116**, 221101 (2016).
[195] B. F. Schutz, Nature (London) **323**, 310 (1986).
[196] D. Eichler, M. Livio, T. Piran, and D. N. Schramm, Nature (London) **340**, 126 (1989).
[197] P. D'Avanzo, J. High Energy Astrophys. **7**, 73 (2015).

B. P. Abbott,[1] R. Abbott,[1] T. D. Abbott,[2] F. Acernese,[3,4] K. Ackley,[5,6] C. Adams,[7] T. Adams,[8] P. Addesso,[9] R. X. Adhikari,[1] V. B. Adya,[10] C. Affeldt,[10] M. Afrough,[11] B. Agarwal,[12] M. Agathos,[13] K. Agatsuma,[14] N. Aggarwal,[15] O. D. Aguiar,[16] L. Aiello,[17,18] A. Ain,[19] P. Ajith,[20] B. Allen,[10,21,22] G. Allen,[12] A. Allocca,[23,24] P. A. Altin,[25] A. Amato,[26] A. Ananyeva,[1] S. B. Anderson,[1] W. G. Anderson,[21] S. V. Angelova,[27] S. Antier,[28] S. Appert,[1] K. Arai,[1] M. C. Araya,[1] J. S. Areeda,[29] N. Arnaud,[28,30] K. G. Arun,[31] S. Ascenzi,[32,33] G. Ashton,[10] M. Ast,[34] S. M. Aston,[7] P. Astone,[35] D. V. Atallah,[36] P. Aufmuth,[22] C. Aulbert,[10] K. AultONeal,[37] C. Austin,[2] A. Avila-Alvarez,[29] S. Babak,[38] P. Bacon,[39] M. K. M. Bader,[14] S. Bae,[40] M. Bailes,[41] P. T. Baker,[42] F. Baldaccini,[43,44] G. Ballardin,[30] S. W. Ballmer,[45] S. Banagiri,[46] J. C. Barayoga,[1] S. E. Barclay,[47] B. C. Barish,[1] D. Barker,[48] K. Barkett,[49] F. Barone,[3,4] B. Barr,[47] L. Barsotti,[15] M. Barsuglia,[39] D. Barta,[50] S. D. Barthelmy,[51] J. Bartlett,[48] I. Bartos,[52,5] R. Bassiri,[53] A. Basti,[23,24] J. C. Batch,[48] M. Bawaj,[54,44] J. C. Bayley,[47] M. Bazzan,[55,56] B. Bécsy,[57] C. Beer,[10] M. Bejger,[58] I. Belahcene,[28] A. S. Bell,[47] B. K. Berger,[1] G. Bergmann,[10] S. Bernuzzi,[59,60] J. J. Bero,[61] C. P. L. Berry,[62] D. Bersanetti,[63] A. Bertolini,[14] J. Betzwieser,[7] S. Bhagwat,[45] R. Bhandare,[64] I. A. Bilenko,[65] G. Billingsley,[1] C. R. Billman,[5] J. Birch,[7] R. Birney,[66] O. Birnholtz,[10] S. Biscans,[1,15] S. Biscoveanu,[67,6] A. Bisht,[22] M. Bitossi,[30,24] C. Biwer,[45] M. A. Bizouard,[28] J. K. Blackburn,[1] J. Blackman,[49] C. D. Blair,[1,68] D. G. Blair,[68] R. M. Blair,[48] S. Bloemen,[69] O. Bock,[10] N. Bode,[10] M. Boer,[70] G. Bogaert,[70] A. Bohe,[38] F. Bondu,[71] E. Bonilla,[53] R. Bonnand,[8] B. A. Boom,[14] R. Bork,[1] V. Boschi,[30,24] S. Bose,[72,19] K. Bossie,[7] Y. Bouffanais,[39] A. Bozzi,[30] C. Bradaschia,[24] P. R. Brady,[21] M. Branchesi,[17,18] J. E. Brau,[73] T. Briant,[74] A. Brillet,[70] M. Brinkmann,[10] V. Brisson,[28] P. Brockill,[21] J. E. Broida,[75] A. F. Brooks,[1] D. A. Brown,[45] D. D. Brown,[76] S. Brunett,[1] C. C. Buchanan,[2] A. Buikema,[15] T. Bulik,[77] H. J. Bulten,[78,14] A. Buonanno,[38,79] D. Buskulic,[8] C. Buy,[39] R. L. Byer,[53] M. Cabero,[10] L. Cadonati,[80] G. Cagnoli,[26,81] C. Cahillane,[1] J. Calderón Bustillo,[80] T. A. Callister,[1] E. Calloni,[82,4] J. B. Camp,[51] M. Canepa,[83,63] P. Canizares,[69] K. C. Cannon,[84] H. Cao,[76] J. Cao,[85] C. D. Capano,[10] E. Capocasa,[39] F. Carbognani,[30] S. Caride,[86] M. F. Carney,[87] G. Carullo,[23,24] J. Casanueva Diaz,[28] C. Casentini,[32,33] S. Caudill,[21,14] M. Cavaglià,[11] F. Cavalier,[28] R. Cavalieri,[30] G. Cella,[24] C. B. Cepeda,[1] P. Cerdá-Durán,[88] G. Cerretani,[23,24] E. Cesarini,[89,33] S. J. Chamberlin,[67] M. Chan,[47] S. Chao,[90] P. Charlton,[91] E. Chase,[92] E. Chassande-Mottin,[39] D. Chatterjee,[21] K. Chatziioannou,[93] B. D. Cheeseboro,[42] H. Y. Chen,[94] X. Chen,[68] Y. Chen,[49] H.-P. Cheng,[5] H. Chia,[5] A. Chincarini,[63] A. Chiummo,[30] T. Chmiel,[87] H. S. Cho,[95] M. Cho,[79] J. H. Chow,[25] N. Christensen,[75,70] Q. Chu,[68] A. J. K. Chua,[13] S. Chua,[74] A. K. W. Chung,[96] S. Chung,[68] G. Ciani,[5,55,56] R. Ciolfi,[97,98] C. E. Cirelli,[53] A. Cirone,[83,63] F. Clara,[48] J. A. Clark,[80] P. Clearwater,[99] F. Cleva,[70] C. Cocchieri,[11] E. Coccia,[17,18] P.-F. Cohadon,[74] D. Cohen,[28] A. Colla,[100,35] C. G. Collette,[101] L. R. Cominsky,[102] M. Constancio Jr.,[16] L. Conti,[56] S. J. Cooper,[62] P. Corban,[7] T. R. Corbitt,[2] I. Cordero-Carrión,[103] K. R. Corley,[52] N. Cornish,[104] A. Corsi,[86] S. Cortese,[30]

C. A. Costa,[16] M. W. Coughlin,[75,1] S. B. Coughlin,[92] J.-P. Coulon,[70] S. T. Countryman,[52] P. Couvares,[1] P. B. Covas,[105] E. E. Cowan,[80] D. M. Coward,[68] M. J. Cowart,[7] D. C. Coyne,[1] R. Coyne,[86] J. D. E. Creighton,[21] T. D. Creighton,[106] J. Cripe,[2] S. G. Crowder,[107] T. J. Cullen,[29,2] A. Cumming,[47] L. Cunningham,[47] E. Cuoco,[30] T. Dal Canton,[51] G. Dálya,[57] S. L. Danilishin,[22,10] S. D'Antonio,[33] K. Danzmann,[22,10] A. Dasgupta,[108] C. F. Da Silva Costa,[5] V. Dattilo,[30] I. Dave,[64] M. Davier,[28] D. Davis,[45] E. J. Daw,[109] B. Day,[80] S. De,[45] D. DeBra,[53] J. Degallaix,[26] M. De Laurentis,[17,4] S. Deléglise,[74] W. Del Pozzo,[62,23,24] N. Demos,[15] T. Denker,[10] T. Dent,[10] R. De Pietri,[59,60] V. Dergachev,[38] R. De Rosa,[82,4] R. T. DeRosa,[7] C. De Rossi,[26,30] R. DeSalvo,[110] O. de Varona,[10] J. Devenson,[27] S. Dhurandhar,[19] M. C. Díaz,[106] T. Dietrich,[38] L. Di Fiore,[4] M. Di Giovanni,[111,98] T. Di Girolamo,[52,82,4] A. Di Lieto,[23,24] S. Di Pace,[100,35] I. Di Palma,[100,35] F. Di Renzo,[23,24] Z. Doctor,[94] V. Dolique,[26] F. Donovan,[15] K. L. Dooley,[11] S. Doravari,[10] I. Dorrington,[36] R. Douglas,[47] M. Dovale Álvarez,[62] T. P. Downes,[21] M. Drago,[10] C. Dreissigacker,[10] J. C. Driggers,[48] Z. Du,[85] M. Ducrot,[8] R. Dudi,[36] P. Dupej,[47] S. E. Dwyer,[48] T. B. Edo,[109] M. C. Edwards,[75] A. Effler,[7] H.-B. Eggenstein,[38,10] P. Ehrens,[1] J. Eichholz,[1] S. S. Eikenberry,[5] R. A. Eisenstein,[15] R. C. Essick,[15] D. Estevez,[8] Z. B. Etienne,[42] T. Etzel,[1] M. Evans,[15] T. M. Evans,[7] M. Factourovich,[52] V. Fafone,[32,33,17] H. Fair,[45] S. Fairhurst,[36] X. Fan,[85] S. Farinon,[63] B. Farr,[94] W. M. Farr,[62] E. J. Fauchon-Jones,[36] M. Favata,[112] M. Fays,[36] C. Fee,[87] H. Fehrmann,[10] J. Feicht,[1] M. M. Fejer,[53] A. Fernandez-Galiana,[15] I. Ferrante,[23,24] E. C. Ferreira,[16] F. Ferrini,[30] F. Fidecaro,[23,24] D. Finstad,[45] I. Fiori,[30] D. Fiorucci,[39] M. Fishbach,[94] R. P. Fisher,[45] M. Fitz-Axen,[46] R. Flaminio,[26,113] M. Fletcher,[47] H. Fong,[93] J. A. Font,[88,114] P. W. F. Forsyth,[25] S. S. Forsyth,[80] J.-D. Fournier,[70] S. Frasca,[100,35] F. Frasconi,[24] Z. Frei,[57] A. Freise,[62] R. Frey,[73] V. Frey,[28] E. M. Fries,[1] P. Fritschel,[15] V. V. Frolov,[7] P. Fulda,[5] M. Fyffe,[7] H. Gabbard,[47] B. U. Gadre,[19] S. M. Gaebel,[62] J. R. Gair,[115] L. Gammaitoni,[43] M. R. Ganija,[76] S. G. Gaonkar,[19] C. Garcia-Quiros,[105] F. Garufi,[82,4] B. Gateley,[48] S. Gaudio,[37] G. Gaur,[116] V. Gayathri,[117] N. Gehrels,[51,†] G. Gemme,[63] E. Genin,[30] A. Gennai,[24] D. George,[12] J. George,[64] L. Gergely,[118] V. Germain,[8] S. Ghonge,[80] Abhirup Ghosh,[20] Archisman Ghosh,[20,14] S. Ghosh,[69,14,21] J. A. Giaime,[2,7] K. D. Giardina,[7] A. Giazotto,[24] K. Gill,[37] L. Glover,[110] E. Goetz,[119] R. Goetz,[5] S. Gomes,[36] B. Goncharov,[6] G. González,[2] J. M. Gonzalez Castro,[23,24] A. Gopakumar,[120] M. L. Gorodetsky,[65] S. E. Gossan,[1] M. Gosselin,[30] R. Gouaty,[8] A. Grado,[121,4] C. Graef,[47] M. Granata,[26] A. Grant,[47] S. Gras,[15] C. Gray,[48] G. Greco,[122,123] A. C. Green,[62] E. M. Gretarsson,[37] P. Groot,[69] H. Grote,[10] S. Grunewald,[38] P. Gruning,[28] G. M. Guidi,[122,123] X. Guo,[85] A. Gupta,[67] M. K. Gupta,[108] K. E. Gushwa,[1] E. K. Gustafson,[1] R. Gustafson,[119] O. Halim,[18,17] B. R. Hall,[72] E. D. Hall,[15] E. Z. Hamilton,[36] G. Hammond,[47] M. Haney,[124] M. M. Hanke,[10] J. Hanks,[48] C. Hanna,[67] M. D. Hannam,[36] O. A. Hannuksela,[96] J. Hanson,[7] T. Hardwick,[2] J. Harms,[17,18] G. M. Harry,[125] I. W. Harry,[38] M. J. Hart,[47] C.-J. Haster,[93] K. Haughian,[47] J. Healy,[61] A. Heidmann,[74] M. C. Heintze,[7] H. Heitmann,[70] P. Hello,[28] G. Hemming,[30] M. Hendry,[47] I. S. Heng,[47] J. Hennig,[47] A. W. Heptonstall,[1] M. Heurs,[10,22] S. Hild,[47] T. Hinderer,[69] W. C. G. Ho,[126] D. Hoak,[30] D. Hofman,[26] K. Holt,[7] D. E. Holz,[94] P. Hopkins,[36] C. Horst,[21] J. Hough,[47] E. A. Houston,[47] E. J. Howell,[68] A. Hreibi,[70] Y. M. Hu,[10] E. A. Huerta,[12] D. Huet,[28] B. Hughey,[37] S. Husa,[105] S. H. Huttner,[47] T. Huynh-Dinh,[7] N. Indik,[10] R. Inta,[86] G. Intini,[100,35] H. N. Isa,[47] J.-M. Isac,[74] M. Isi,[1] B. R. Iyer,[20] K. Izumi,[48] T. Jacqmin,[74] K. Jani,[80] P. Jaranowski,[127] S. Jawahar,[66] F. Jiménez-Forteza,[105] W. W. Johnson,[2] N. K. Johnson-McDaniel,[13] D. I. Jones,[126] R. Jones,[47] R. J. G. Jonker,[14] L. Ju,[68] J. Junker,[10] C. V. Kalaghatgi,[36] V. Kalogera,[92] B. Kamai,[1] S. Kandhasamy,[7] G. Kang,[40] J. B. Kanner,[1] S. J. Kapadia,[21] S. Karki,[73] K. S. Karvinen,[10] M. Kasprzack,[2] W. Kastaun,[10] M. Katolik,[12] E. Katsavounidis,[15] W. Katzman,[7] S. Kaufer,[22] K. Kawabe,[48] F. Kéfélian,[70] D. Keitel,[47] A. J. Kemball,[12] R. Kennedy,[109] C. Kent,[36] J. S. Key,[128] F. Y. Khalili,[65] I. Khan,[17,33] S. Khan,[10] Z. Khan,[108] E. A. Khazanov,[129] N. Kijbunchoo,[25] Chunglee Kim,[130] J. C. Kim,[131] K. Kim,[96] W. Kim,[76] W. S. Kim,[132] Y.-M. Kim,[95] S. J. Kimbrell,[80] E. J. King,[76] P. J. King,[48] M. Kinley-Hanlon,[125] R. Kirchhoff,[10] J. S. Kissel,[48] L. Kleybolte,[34] S. Klimenko,[5] T. D. Knowles,[42] P. Koch,[10] S. M. Koehlenbeck,[10] S. Koley,[14] V. Kondrashov,[1] A. Kontos,[15] M. Korobko,[34] W. Z. Korth,[1] I. Kowalska,[77] D. B. Kozak,[1] C. Krämer,[10] V. Kringel,[10] B. Krishnan,[10] A. Królak,[133,134] G. Kuehn,[10] P. Kumar,[93] R. Kumar,[108] S. Kumar,[20] L. Kuo,[90] A. Kutynia,[133] S. Kwang,[21] B. D. Lackey,[38] K. H. Lai,[96] M. Landry,[48] R. N. Lang,[135] J. Lange,[61] B. Lantz,[53] R. K. Lanza,[15] S. L. Larson,[92] A. Lartaux-Vollard,[28] P. D. Lasky,[6] M. Laxen,[7] A. Lazzarini,[1] C. Lazzaro,[56] P. Leaci,[100,35] S. Leavey,[47] C. H. Lee,[95] H. K. Lee,[136] H. M. Lee,[137] H. W. Lee,[131] K. Lee,[47] J. Lehmann,[10] A. Lenon,[42] E. Leon,[29] M. Leonardi,[111,98] N. Leroy,[28] N. Letendre,[8] Y. Levin,[6] T. G. F. Li,[96] S. D. Linker,[110] T. B. Littenberg,[138] J. Liu,[68] X. Liu,[21] R. K. L. Lo,[96] N. A. Lockerbie,[66] L. T. London,[36] J. E. Lord,[45] M. Lorenzini,[17,18] V. Loriette,[139] M. Lormand,[7] G. Losurdo,[24] J. D. Lough,[10] C. O. Lousto,[61] G. Lovelace,[29] H. Lück,[22,10] D. Lumaca,[32,33] A. P. Lundgren,[10] R. Lynch,[15] Y. Ma,[49] R. Macas,[36] S. Macfoy,[27] B. Machenschalk,[10] M. MacInnis,[15] D. M. Macleod,[36] I. Magaña Hernandez,[21] F. Magaña-Sandoval,[45] L. Magaña Zertuche,[45] R. M. Magee,[67] E. Majorana,[35] I. Maksimovic,[139] N. Man,[70] V. Mandic,[46] V. Mangano,[47]

G. L. Mansell,[25] M. Manske,[21,25] M. Mantovani,[30] F. Marchesoni,[54,44] F. Marion,[8] S. Márka,[52] Z. Márka,[52] C. Markakis,[12] A. S. Markosyan,[53] A. Markowitz,[1] E. Maros,[1] A. Marquina,[103] P. Marsh,[128] F. Martelli,[122,123] L. Martellini,[70] I. W. Martin,[47] R. M. Martin,[112] D. V. Martynov,[15] J. N. Marx,[1] K. Mason,[15] E. Massera,[109] A. Masserot,[8] T. J. Massinger,[1] M. Masso-Reid,[47] S. Mastrogiovanni,[100,35] A. Matas,[46] F. Matichard,[1,15] L. Matone,[52] N. Mavalvala,[15] N. Mazumder,[72] R. McCarthy,[48] D. E. McClelland,[25] S. McCormick,[7] L. McCuller,[15] S. C. McGuire,[140] G. McIntyre,[1] J. McIver,[1] D. J. McManus,[25] L. McNeill,[6] T. McRae,[25] S. T. McWilliams,[42] D. Meacher,[67] G. D. Meadors,[38,10] M. Mehmet,[10] J. Meidam,[14] E. Mejuto-Villa,[9] A. Melatos,[99] G. Mendell,[48] R. A. Mercer,[21] E. L. Merilh,[48] M. Merzougui,[70] S. Meshkov,[1] C. Messenger,[47] C. Messick,[67] R. Metzdorff,[74] P. M. Meyers,[46] H. Miao,[62] C. Michel,[26] H. Middleton,[62] E. E. Mikhailov,[141] L. Milano,[82,4] A. L. Miller,[5,100,35] B. B. Miller,[92] J. Miller,[15] M. Millhouse,[104] M. C. Milovich-Goff,[110] O. Minazzoli,[70,142] Y. Minenkov,[33] J. Ming,[38] C. Mishra,[143] S. Mitra,[19] V. P. Mitrofanov,[65] G. Mitselmakher,[5] R. Mittleman,[15] D. Moffa,[87] A. Moggi,[24] K. Mogushi,[11] M. Mohan,[30] S. R. P. Mohapatra,[15] I. Molina,[29] M. Montani,[122,123] C. J. Moore,[13] D. Moraru,[48] G. Moreno,[48] S. Morisaki,[84] S. R. Morriss,[106] B. Mours,[8] C. M. Mow-Lowry,[62] G. Mueller,[5] A. W. Muir,[36] Arunava Mukherjee,[10] D. Mukherjee,[21] S. Mukherjee,[106] N. Mukund,[19] A. Mullavey,[7] J. Munch,[76] E. A. Muñiz,[45] M. Muratore,[37] P. G. Murray,[47] A. Nagar,[89,144,145] K. Napier,[80] I. Nardecchia,[32,33] L. Naticchioni,[100,35] R. K. Nayak,[146] J. Neilson,[110] G. Nelemans,[69,14] T. J. N. Nelson,[7] M. Nery,[10] A. Neunzert,[119] L. Nevin,[1] J. M. Newport,[125] G. Newton,[47,‡] K. K. Y. Ng,[96] P. Nguyen,[73] T. T. Nguyen,[25] D. Nichols,[69] A. B. Nielsen,[10] S. Nissanke,[69,14] A. Nitz,[10] A. Noack,[10] F. Nocera,[30] D. Nolting,[7] C. North,[36] L. K. Nuttall,[36] J. Oberling,[48] G. D. O'Dea,[110] G. H. Ogin,[147] J. J. Oh,[132] S. H. Oh,[132] F. Ohme,[10] M. A. Okada,[16] M. Oliver,[105] P. Oppermann,[10] Richard J. Oram,[7] B. O'Reilly,[7] R. Ormiston,[46] L. F. Ortega,[5] R. O'Shaughnessy,[61] S. Ossokine,[38] D. J. Ottaway,[76] H. Overmier,[7] B. J. Owen,[86] A. E. Pace,[67] J. Page,[138] M. A. Page,[68] A. Pai,[117,148] S. A. Pai,[64] J. R. Palamos,[73] O. Palashov,[129] C. Palomba,[35] A. Pal-Singh,[34] Howard Pan,[90] Huang-Wei Pan,[90] B. Pang,[49] P. T. H. Pang,[96] C. Pankow,[92] F. Pannarale,[36] B. C. Pant,[64] F. Paoletti,[24] A. Paoli,[30] M. A. Papa,[38,21,10] A. Parida,[19] W. Parker,[7] D. Pascucci,[47] A. Pasqualetti,[30] R. Passaquieti,[23,24] D. Passuello,[24] M. Patil,[134] B. Patricelli,[149,24] B. L. Pearlstone,[47] M. Pedraza,[1] R. Pedurand,[26,150] L. Pekowsky,[45] A. Pele,[7] S. Penn,[151] C. J. Perez,[48] A. Perreca,[1,111,98] L. M. Perri,[92] H. P. Pfeiffer,[93,38] M. Phelps,[47] O. J. Piccinni,[100,35] M. Pichot,[70] F. Piergiovanni,[122,123] V. Pierro,[9] G. Pillant,[30] L. Pinard,[26] I. M. Pinto,[9] M. Pirello,[48] M. Pitkin,[47] M. Poe,[21] R. Poggiani,[23,24] P. Popolizio,[30] E. K. Porter,[39] A. Post,[10] J. Powell,[47,41] J. Prasad,[19] J. W. W. Pratt,[37] G. Pratten,[105] V. Predoi,[36] T. Prestegard,[21] M. Prijatelj,[10] M. Principe,[9] S. Privitera,[38] R. Prix,[10] G. A. Prodi,[111,98] L. G. Prokhorov,[65] O. Puncken,[10] M. Punturo,[44] P. Puppo,[35] M. Pürrer,[38] H. Qi,[21] V. Quetschke,[106] E. A. Quintero,[1] R. Quitzow-James,[73] F. J. Raab,[48] D. S. Rabeling,[25] H. Radkins,[48] P. Raffai,[57] S. Raja,[64] C. Rajan,[64] B. Rajbhandari,[86] M. Rakhmanov,[106] K. E. Ramirez,[106] A. Ramos-Buades,[105] P. Rapagnani,[100,35] V. Raymond,[38] M. Razzano,[23,24] J. Read,[29] T. Regimbau,[70] L. Rei,[63] S. Reid,[66] D. H. Reitze,[1,5] W. Ren,[12] S. D. Reyes,[45] F. Ricci,[100,35] P. M. Ricker,[12] S. Rieger,[10] K. Riles,[119] M. Rizzo,[61] N. A. Robertson,[1,47] R. Robie,[47] F. Robinet,[28] A. Rocchi,[33] L. Rolland,[8] J. G. Rollins,[1] V. J. Roma,[73] J. D. Romano,[106] R. Romano,[3,4] C. L. Romel,[48] J. H. Romie,[7] D. Rosińska,[152,58] M. P. Ross,[153] S. Rowan,[47] A. Rüdiger,[10] P. Ruggi,[30] G. Rutins,[27] K. Ryan,[48] S. Sachdev,[1] T. Sadecki,[48] L. Sadeghian,[21] M. Sakellariadou,[154] L. Salconi,[30] M. Saleem,[117] F. Salemi,[10] A. Samajdar,[146] L. Sammut,[6] L. M. Sampson,[92] E. J. Sanchez,[1] L. E. Sanchez,[1] N. Sanchis-Gual,[88] V. Sandberg,[48] J. R. Sanders,[45] B. Sassolas,[26] B. S. Sathyaprakash,[67,36] P. R. Saulson,[45] O. Sauter,[119] R. L. Savage,[48] A. Sawadsky,[34] P. Schale,[73] M. Scheel,[49] J. Scheuer,[92] J. Schmidt,[10] P. Schmidt,[1,69] R. Schnabel,[34] R. M. S. Schofield,[73] A. Schönbeck,[34] E. Schreiber,[10] D. Schuette,[10,22] B. W. Schulte,[10] B. F. Schutz,[36,10] S. G. Schwalbe,[37] J. Scott,[47] S. M. Scott,[25] E. Seidel,[12] D. Sellers,[7] A. S. Sengupta,[155] D. Sentenac,[30] V. Sequino,[32,33,17] A. Sergeev,[129] D. A. Shaddock,[25] T. J. Shaffer,[48] A. A. Shah,[138] M. S. Shahriar,[92] M. B. Shaner,[110] L. Shao,[38] B. Shapiro,[53] P. Shawhan,[79] A. Sheperd,[21] D. H. Shoemaker,[15] D. M. Shoemaker,[80] K. Siellez,[80] X. Siemens,[21] M. Sieniawska,[58] D. Sigg,[48] A. D. Silva,[16] L. P. Singer,[51] A. Singh,[38,10,22] A. Singhal,[17,35] A. M. Sintes,[105] B. J. J. Slagmolen,[25] B. Smith,[7] J. R. Smith,[29] R. J. E. Smith,[1,6] S. Somala,[156] E. J. Son,[132] J. A. Sonnenberg,[21] B. Sorazu,[47] F. Sorrentino,[63] T. Souradeep,[19] A. P. Spencer,[47] A. K. Srivastava,[108] K. Staats,[37] A. Staley,[52] M. Steinke,[10] J. Steinlechner,[34,47] S. Steinlechner,[34] D. Steinmeyer,[10] S. P. Stevenson,[62,41] R. Stone,[106] D. J. Stops,[62] K. A. Strain,[47] G. Stratta,[122,123] S. E. Strigin,[65] A. Strunk,[48] R. Sturani,[157] A. L. Stuver,[7] T. Z. Summerscales,[158] L. Sun,[99] S. Sunil,[108] J. Suresh,[19] P. J. Sutton,[36] B. L. Swinkels,[30] M. J. Szczepańczyk,[37] M. Tacca,[14] S. C. Tait,[47] C. Talbot,[6] D. Talukder,[73] D. B. Tanner,[5] M. Tápai,[118] A. Taracchini,[38] J. D. Tasson,[75] J. A. Taylor,[138] R. Taylor,[1] S. V. Tewari,[151] T. Theeg,[10] F. Thies,[10] E. G. Thomas,[62] M. Thomas,[7] P. Thomas,[48] K. A. Thorne,[7] K. S. Thorne,[49] E. Thrane,[6] S. Tiwari,[17,98] V. Tiwari,[36] K. V. Tokmakov,[66] K. Toland,[47] M. Tonelli,[23,24] Z. Tornasi,[47] A. Torres-Forné,[88] C. I. Torrie,[1] D. Töyrä,[62] F. Travasso,[30,44] G. Traylor,[7] J. Trinastic,[5] M. C. Tringali,[111,98]

L. Trozzo,[159,24] K. W. Tsang,[14] M. Tse,[15] R. Tso,[1] L. Tsukada,[84] D. Tsuna,[84] D. Tuyenbayev,[106] K. Ueno,[21] D. Ugolini,[160] C. S. Unnikrishnan,[120] A. L. Urban,[1] S. A. Usman,[36] H. Vahlbruch,[22] G. Vajente,[1] G. Valdes,[2] M. Vallisneri,[49] N. van Bakel,[14] M. van Beuzekom,[14] J. F. J. van den Brand,[78,14] C. Van Den Broeck,[14] D. C. Vander-Hyde,[45] L. van der Schaaf,[14] J. V. van Heijningen,[14] A. A. van Veggel,[47] M. Vardaro,[55,56] V. Varma,[49] S. Vass,[1] M. Vasúth,[50] A. Vecchio,[62] G. Vedovato,[56] J. Veitch,[47] P. J. Veitch,[76] K. Venkateswara,[153] G. Venugopalan,[1] D. Verkindt,[8] F. Vetrano,[122,123] A. Viceré,[122,123] A. D. Viets,[21] S. Vinciguerra,[62] D. J. Vine,[27] J.-Y. Vinet,[70] S. Vitale,[15] T. Vo,[45] H. Vocca,[43,44] C. Vorvick,[48] S. P. Vyatchanin,[65] A. R. Wade,[1] L. E. Wade,[87] M. Wade,[87] R. Walet,[14] M. Walker,[29] L. Wallace,[1] S. Walsh,[38,10,21] G. Wang,[17,123] H. Wang,[62] J. Z. Wang,[67] W. H. Wang,[106] Y. F. Wang,[96] R. L. Ward,[25] J. Warner,[48] M. Was,[8] J. Watchi,[101] B. Weaver,[48] L.-W. Wei,[10,22] M. Weinert,[10] A. J. Weinstein,[1] R. Weiss,[15] L. Wen,[68] E. K. Wessel,[12] P. Weßels,[10] J. Westerweck,[10] T. Westphal,[10] K. Wette,[25] J. T. Whelan,[61] S. E. Whitcomb,[1] B. F. Whiting,[5] C. Whittle,[6] D. Wilken,[10] D. Williams,[47] R. D. Williams,[1] A. R. Williamson,[69] J. L. Willis,[1,161] B. Willke,[22,10] M. H. Wimmer,[10] W. Winkler,[10] C. C. Wipf,[1] H. Wittel,[10,22] G. Woan,[47] J. Woehler,[10] J. Wofford,[61] K. W. K. Wong,[96] J. Worden,[48] J. L. Wright,[47] D. S. Wu,[10] D. M. Wysocki,[61] S. Xiao,[1] H. Yamamoto,[1] C. C. Yancey,[79] L. Yang,[162] M. J. Yap,[25] M. Yazback,[5] Hang Yu,[15] Haocun Yu,[15] M. Yvert,[8] A. Zadrożny,[133] M. Zanolin,[37] T. Zelenova,[30] J.-P. Zendri,[56] M. Zevin,[92] L. Zhang,[1] M. Zhang,[141] T. Zhang,[47] Y.-H. Zhang,[61] C. Zhao,[68] M. Zhou,[92] Z. Zhou,[92] S. J. Zhu,[38,10] X. J. Zhu,[6] A. B. Zimmerman,[93] M. E. Zucker,[1,15] and J. Zweizig[1]

(LIGO Scientific Collaboration and Virgo Collaboration)

[1]*LIGO, California Institute of Technology, Pasadena, California 91125, USA*
[2]*Louisiana State University, Baton Rouge, Louisiana 70803, USA*
[3]*Università di Salerno, Fisciano, I-84084 Salerno, Italy*
[4]*INFN, Sezione di Napoli, Complesso Universitario di Monte S.Angelo, I-80126 Napoli, Italy*
[5]*University of Florida, Gainesville, Florida 32611, USA*
[6]*OzGrav, School of Physics & Astronomy, Monash University, Clayton 3800, Victoria, Australia*
[7]*LIGO Livingston Observatory, Livingston, Louisiana 70754, USA*
[8]*Laboratoire d'Annecy-le-Vieux de Physique des Particules (LAPP), Université Savoie Mont Blanc, CNRS/IN2P3, F-74941 Annecy, France*
[9]*University of Sannio at Benevento, I-82100 Benevento, Italy and INFN, Sezione di Napoli, I-80100 Napoli, Italy*
[10]*Max Planck Institute for Gravitational Physics (Albert Einstein Institute), D-30167 Hannover, Germany*
[11]*The University of Mississippi, University, Mississippi 38677, USA*
[12]*NCSA, University of Illinois at Urbana-Champaign, Urbana, Illinois 61801, USA*
[13]*University of Cambridge, Cambridge CB2 1TN, United Kingdom*
[14]*Nikhef, Science Park, 1098 XG Amsterdam, Netherlands*
[15]*LIGO, Massachusetts Institute of Technology, Cambridge, Massachusetts 02139, USA*
[16]*Instituto Nacional de Pesquisas Espaciais, 12227-010 São José dos Campos, São Paulo, Brazil*
[17]*Gran Sasso Science Institute (GSSI), I-67100 L'Aquila, Italy*
[18]*INFN, Laboratori Nazionali del Gran Sasso, I-67100 Assergi, Italy*
[19]*Inter-University Centre for Astronomy and Astrophysics, Pune 411007, India*
[20]*International Centre for Theoretical Sciences, Tata Institute of Fundamental Research, Bengaluru 560089, India*
[21]*University of Wisconsin-Milwaukee, Milwaukee, Wisconsin 53201, USA*
[22]*Leibniz Universität Hannover, D-30167 Hannover, Germany*
[23]*Università di Pisa, I-56127 Pisa, Italy*
[24]*INFN, Sezione di Pisa, I-56127 Pisa, Italy*
[25]*OzGrav, Australian National University, Canberra, Australian Capital Territory 0200, Australia*
[26]*Laboratoire des Matériaux Avancés (LMA), CNRS/IN2P3, F-69622 Villeurbanne, France*
[27]*SUPA, University of the West of Scotland, Paisley PA1 2BE, United Kingdom*
[28]*LAL, Univ. Paris-Sud, CNRS/IN2P3, Université Paris-Saclay, F-91898 Orsay, France*
[29]*California State University Fullerton, Fullerton, California 92831, USA*
[30]*European Gravitational Observatory (EGO), I-56021 Cascina, Pisa, Italy*
[31]*Chennai Mathematical Institute, Chennai 603103, India*
[32]*Università di Roma Tor Vergata, I-00133 Roma, Italy*
[33]*INFN, Sezione di Roma Tor Vergata, I-00133 Roma, Italy*
[34]*Universität Hamburg, D-22761 Hamburg, Germany*
[35]*INFN, Sezione di Roma, I-00185 Roma, Italy*

[36]*Cardiff University, Cardiff CF24 3AA, United Kingdom*
[37]*Embry-Riddle Aeronautical University, Prescott, Arizona 86301, USA*
[38]*Max Planck Institute for Gravitationalphysik (Albert Einstein Institute), D-14476 Potsdam-Golm, Germany*
[39]*APC, AstroParticule et Cosmologie, Université Paris Diderot, CNRS/IN2P3, CEA/Irfu, Observatoire de Paris, Sorbonne Paris Cité, F-75205 Paris Cedex 13, France*
[40]*Korea Institute of Science and Technology Information, Daejeon 34141, Korea*
[41]*OzGrav, Swinburne University of Technology, Hawthorn VIC 3122, Australia*
[42]*West Virginia University, Morgantown, West Virginia 26506, USA*
[43]*Università di Perugia, I-06123 Perugia, Italy*
[44]*INFN, Sezione di Perugia, I-06123 Perugia, Italy*
[45]*Syracuse University, Syracuse, New York 13244, USA*
[46]*University of Minnesota, Minneapolis, Minnesota 55455, USA*
[47]*SUPA, University of Glasgow, Glasgow G12 8QQ, United Kingdom*
[48]*LIGO Hanford Observatory, Richland, Washington 99352, USA*
[49]*Caltech CaRT, Pasadena, California 91125, USA*
[50]*Wigner RCP, RMKI, H-1121 Budapest, Konkoly Thege Miklós út 29-33, Hungary*
[51]*NASA Goddard Space Flight Center, Greenbelt, Maryland 20771, USA*
[52]*Columbia University, New York, New York 10027, USA*
[53]*Stanford University, Stanford, California 94305, USA*
[54]*Università di Camerino, Dipartimento di Fisica, I-62032 Camerino, Italy*
[55]*Università di Padova, Dipartimento di Fisica e Astronomia, I-35131 Padova, Italy*
[56]*INFN, Sezione di Padova, I-35131 Padova, Italy*
[57]*Institute of Physics, Eötvös University, Pázmány P. s. 1/A, Budapest 1117, Hungary*
[58]*Nicolaus Copernicus Astronomical Center, Polish Academy of Sciences, 00-716, Warsaw, Poland*
[59]*Dipartimento di Scienze Matematiche, Fisiche e Informatiche, Università di Parma, I-43124 Parma, Italy*
[60]*INFN, Sezione di Milano Bicocca, Gruppo Collegato di Parma, I-43124 Parma, Italy*
[61]*Rochester Institute of Technology, Rochester, New York 14623, USA*
[62]*University of Birmingham, Birmingham B15 2TT, United Kingdom*
[63]*INFN, Sezione di Genova, I-16146 Genova, Italy*
[64]*RRCAT, Indore MP 452013, India*
[65]*Faculty of Physics, Lomonosov Moscow State University, Moscow 119991, Russia*
[66]*SUPA, University of Strathclyde, Glasgow G1 1XQ, United Kingdom*
[67]*The Pennsylvania State University, University Park, Pennsylvania 16802, USA*
[68]*OzGrav, University of Western Australia, Crawley, Western Australia 6009, Australia*
[69]*Department of Astrophysics/IMAPP, Radboud University Nijmegen, P.O. Box 9010, 6500 GL Nijmegen, Netherlands*
[70]*Artemis, Université Côte d'Azur, Observatoire Côte d'Azur, CNRS, CS 34229, F-06304 Nice Cedex 4, France*
[71]*Institut FOTON, CNRS, Université de Rennes 1, F-35042 Rennes, France*
[72]*Washington State University, Pullman, Washington 99164, USA*
[73]*University of Oregon, Eugene, Oregon 97403, USA*
[74]*Laboratoire Kastler Brossel, UPMC-Sorbonne Universités, CNRS, ENS-PSL Research University, Collège de France, F-75005 Paris, France*
[75]*Carleton College, Northfield, Minnesota 55057, USA*
[76]*OzGrav, University of Adelaide, Adelaide, South Australia 5005, Australia*
[77]*Astronomical Observatory Warsaw University, 00-478 Warsaw, Poland*
[78]*VU University Amsterdam, 1081 HV Amsterdam, Netherlands*
[79]*University of Maryland, College Park, Maryland 20742, USA*
[80]*Center for Relativistic Astrophysics, Georgia Institute of Technology, Atlanta, Georgia 30332, USA*
[81]*Université Claude Bernard Lyon 1, F-69622 Villeurbanne, France*
[82]*Università di Napoli "Federico II", Complesso Universitario di Monte S.Angelo, I-80126 Napoli, Italy*
[83]*Dipartimento di Fisica, Università degli Studi di Genova, I-16146 Genova, Italy*
[84]*RESCEU, University of Tokyo, Tokyo, 113-0033, Japan*
[85]*Tsinghua University, Beijing 100084, China*
[86]*Texas Tech University, Lubbock, Texas 79409, USA*
[87]*Kenyon College, Gambier, Ohio 43022, USA*
[88]*Departamento de Astronomía y Astrofísica, Universitat de València, E-46100 Burjassot, València, Spain*
[89]*Museo Storico della Fisica e Centro Studi e Ricerche Enrico Fermi, I-00184 Roma, Italy*
[90]*National Tsing Hua University, Hsinchu City, 30013 Taiwan, Republic of China*
[91]*Charles Sturt University, Wagga Wagga, New South Wales 2678, Australia*
[92]*Center for Interdisciplinary Exploration & Research in Astrophysics (CIERA), Northwestern University, Evanston, Illinois 60208, USA*

[93]*Canadian Institute for Theoretical Astrophysics, University of Toronto, Toronto, Ontario M5S 3H8, Canada*
[94]*University of Chicago, Chicago, Illinois 60637, USA*
[95]*Pusan National University, Busan 46241, Korea*
[96]*The Chinese University of Hong Kong, Shatin, NT, Hong Kong*
[97]*INAF, Osservatorio Astronomico di Padova, I-35122 Padova, Italy*
[98]*INFN, Trento Institute for Fundamental Physics and Applications, I-38123 Povo, Trento, Italy*
[99]*OzGrav, University of Melbourne, Parkville, Victoria 3010, Australia*
[100]*Università di Roma "La Sapienza", I-00185 Roma, Italy*
[101]*Université Libre de Bruxelles, Brussels 1050, Belgium*
[102]*Sonoma State University, Rohnert Park, California 94928, USA*
[103]*Departamento de Matemáticas, Universitat de València, E-46100 Burjassot, València, Spain*
[104]*Montana State University, Bozeman, Montana 59717, USA*
[105]*Universitat de les Illes Balears, IAC3—IEEC, E-07122 Palma de Mallorca, Spain*
[106]*The University of Texas Rio Grande Valley, Brownsville, Texas 78520, USA*
[107]*Bellevue College, Bellevue, Washington 98007, USA*
[108]*Institute for Plasma Research, Bhat, Gandhinagar 382428, India*
[109]*The University of Sheffield, Sheffield S10 2TN, United Kingdom*
[110]*California State University, Los Angeles, 5151 State University Drive, Los Angeles, California 90032, USA*
[111]*Università di Trento, Dipartimento di Fisica, I-38123 Povo, Trento, Italy*
[112]*Montclair State University, Montclair, New Jersey 07043, USA*
[113]*National Astronomical Observatory of Japan, 2-21-1 Osawa, Mitaka, Tokyo 181-8588, Japan*
[114]*Observatori Astronòmic, Universitat de València, E-46980 Paterna, València, Spain*
[115]*School of Mathematics, University of Edinburgh, Edinburgh EH9 3FD, United Kingdom*
[116]*University and Institute of Advanced Research, Koba Institutional Area, Gandhinagar Gujarat 382007, India*
[117]*IISER-TVM, CET Campus, Trivandrum Kerala 695016, India*
[118]*University of Szeged, Dóm tér 9, Szeged 6720, Hungary*
[119]*University of Michigan, Ann Arbor, Michigan 48109, USA*
[120]*Tata Institute of Fundamental Research, Mumbai 400005, India*
[121]*INAF, Osservatorio Astronomico di Capodimonte, I-80131, Napoli, Italy*
[122]*Università degli Studi di Urbino "Carlo Bo" I-61029 Urbino, Italy*
[123]*INFN, Sezione di Firenze, I-50019 Sesto Fiorentino, Firenze, Italy*
[124]*Physik-Institut, University of Zurich, Winterthurerstrasse 190, 8057 Zurich, Switzerland*
[125]*American University, Washington, DC 20016, USA*
[126]*University of Southampton, Southampton SO17 1BJ, United Kingdom*
[127]*University of Biał ystok, 15-424 Biał ystok, Poland*
[128]*University of Washington Bothell, 18115 Campus Way NE, Bothell, Washington 98011, USA*
[129]*Institute of Applied Physics, Nizhny Novgorod, 603950, Russia*
[130]*Korea Astronomy and Space Science Institute, Daejeon 34055, Korea*
[131]*Inje University Gimhae, South Gyeongsang 50834, Korea*
[132]*National Institute for Mathematical Sciences, Daejeon 34047, Korea*
[133]*NCBJ, 05-400 Świerk-Otwock, Poland*
[134]*Institute of Mathematics, Polish Academy of Sciences, 00656 Warsaw, Poland*
[135]*Hillsdale College, Hillsdale, Michigan 49242, USA*
[136]*Hanyang University, Seoul 04763, Korea*
[137]*Seoul National University, Seoul 08826, Korea*
[138]*NASA Marshall Space Flight Center, Huntsville, Alabama 35811, USA*
[139]*ESPCI, CNRS, F-75005 Paris, France*
[140]*Southern University and A&M College, Baton Rouge, Louisiana 70813, USA*
[141]*College of William and Mary, Williamsburg, Virginia 23187, USA*
[142]*Centre Scientifique de Monaco, 8 quai Antoine Ier, MC-98000, Monaco*
[143]*Indian Institute of Technology Madras, Chennai 600036, India*
[144]*INFN Sezione di Torino, I-10125 Torino, Italy*
[145]*Institut des Hautes Etudes Scientifiques, F-91440 Bures-sur-Yvette, France*
[146]*IISER-Kolkata, Mohanpur, West Bengal 741252, India*
[147]*Whitman College, 345 Boyer Avenue, Walla Walla, Washington 99362 USA*
[148]*Indian Institute of Technology Bombay, Powai, Mumbai, Maharashtra 400076, India*
[149]*Scuola Normale Superiore, Piazza dei Cavalieri 7, I-56126 Pisa, Italy*
[150]*Université de Lyon, F-69361 Lyon, France*
[151]*Hobart and William Smith Colleges, Geneva, New York 14456, USA*
[152]*Janusz Gil Institute of Astronomy, University of Zielona Góra, 65-265 Zielona Góra, Poland*

[153]*University of Washington, Seattle, Washington 98195, USA*
[154]*King's College London, University of London, London WC2R 2LS, United Kingdom*
[155]*Indian Institute of Technology, Gandhinagar Ahmedabad Gujarat 382424, India*
[156]*Indian Institute of Technology Hyderabad, Sangareddy, Khandi, Telangana 502285, India*
[157]*International Institute of Physics, Universidade Federal do Rio Grande do Norte, Natal RN 59078-970, Brazil*
[158]*Andrews University, Berrien Springs, Michigan 49104, USA*
[159]*Università di Siena, I-53100 Siena, Italy*
[160]*Trinity University, San Antonio, Texas 78212, USA*
[161]*Abilene Christian University, Abilene, Texas 79699, USA*
[162]*Colorado State University, Fort Collins, Colorado 80523, USA*

[†]Deceased, February 2017.
[‡]Deceased, December 2016.